\documentclass{article}

\usepackage{natbib}
\usepackage{epubtk}
\usepackage{graphicx}
\usepackage{booktabs}
\graphicspath{{figs/},{upload1/},{../}}

\newcommand{\dv}[2]{\frac{{\mathrm d}#1}{{\mathrm d}#2}}
\newcommand{\pdv}[2]{\frac{\partial #1}{\partial #2}}
\newcommand{\vc}[1]{\mathbf{#1}}
\newcommand{\sgn}{{\mathrm{sgn}\,}}
\newcommand{\Rmax}{R_{\mathrm{max}}}
\newcommand{\Rmin}{R_{\mathrm{min}}}
\newcommand{\tmax}{t_{\mathrm{max}}}
\newcommand{\tmin}{t_{\mathrm{min}}}
\def\la{\mathrel{\mathchoice {\vcenter{\offinterlineskip\halign{\hfil
 $\displaystyle##$\hfil\cr<\cr\sim\cr}}}
 {\vcenter{\offinterlineskip\halign{\hfil$\textstyle##$\hfil\cr
 <\cr\sim\cr}}}
 {\vcenter{\offinterlineskip\halign{\hfil$\scriptstyle##$\hfil\cr
 <\cr\sim\cr}}}
 {\vcenter{\offinterlineskip\halign{\hfil$\scriptscriptstyle##$\hfil\cr
 <\cr\sim\cr}}}}}

\newenvironment{descr}{\begin{description}}{\end{description}}
\newenvironment{nquote}{\bigskip\strut\hfill\begin{minipage}{0.8\textwidth}}{\end{minipage}\bigskip}

\newcommand{\bek}{\strut\hskip 2em}

\begin{document}

\title{Solar Cycle Prediction}

\author{\epubtkAuthorData{Krist\'of Petrovay}{%
E{\"o}tv{\"o}s University, Department of Astronomy \\
Budapest, Hungary}{%
K.Petrovay@astro.elte.hu}{%
http://astro.elte.hu/\~kris}%
}

\date{}
\maketitle

\begin{abstract} 
A review of solar cycle prediction methods and their performance is given,
including forecasts for cycle~24. The review focuses on those aspects of the
solar cycle prediction problem that have a bearing on dynamo theory. The scope
of the review is further restricted to the issue of predicting the amplitude
(and optionally the epoch) of an upcoming solar maximum no later than right
after the start of the given cycle. 

Prediction methods form three main groups. \emph{Precursor methods} rely on the
value of some measure of solar activity or magnetism at a specified time to
predict the amplitude of the following solar maximum. Their implicit assumption
is that each numbered solar cycle is a consistent unit in itself, while solar
activity seems to consist of a series of much less tightly intercorrelated
individual cycles. \emph{Extrapolation methods,} in contrast, are based on the
premise that the physical process giving rise to the sunspot number record is
statistically  homogeneous, i.e., the mathematical regularities underlying its
variations are the same at any point of time, and therefore it lends itself to
analysis and forecasting by time series methods. Finally, instead of an analysis
of observational data alone, \emph{model based predictions} use physically (more
or less) consistent dynamo models in their attempts to predict solar activity.

In their overall performance during the course of the last few solar cycles,
precursor methods have clearly been superior to extrapolation methods.
Nevertheless, most precursor methods overpredicted cycle~23, while some
extrapolation methods may still be worth further study. Model based forecasts
have not yet have had a chance to prove their skills. One method that has
yielded predictions consistently in the right range during the past few solar
cycles is that of K.~Schatten \textit{et al.}, whose approach is
mainly based on the polar field precursor.

The incipient cycle~24 will probably mark the end of the Modern Maximum, with
the Sun switching to a state of less strong activity. It will therefore be an
important testbed for cycle prediction methods, and, by inference, for our
understanding of the solar dynamo.
\end{abstract}

\epubtkKeywords{solar cycle, solar dynamo}


\newpage

\section{Introduction}
\label{section:introduction}

Solar cycle prediction is an extremely extensive topic, covering a very wide
variety of proposed prediction methods and prediction attempts on many different
timescales, ranging from short term (month--year) forecasts of the runoff of the
ongoing solar cycle to predictions of long term changes in solar activity on 
centennial or even millennial scales. As early as 1963, Vitinsky published a
whole monograph on the subject, later updated and extended
\citep{Vitinsky:book1, Vitinsky:book2}. More recent overviews of
the field or aspects of it include \cite{Hathaway:prediction.review},
\cite{Kane:pred23rev},  and \cite{Pesnell}. In order to narrow down the scope of
the present review, we constrain our field of interest in two important
respects.

Firstly, instead of attempting to give a general review of all prediction
methods suggested or citing all the papers with forecasts, here we will focus on
those aspects of the solar cycle prediction problem that have a bearing on
dynamo theory. We will thus discuss in more detail empirical methods that,
independently of their success rate, have the potential of shedding some light
on the physical mechanism underlying the solar cycle, as well as the prediction
attempts based on solar dynamo models.

Secondly, we will here only be concerned with the issue of \textit{predicting the
amplitude} (and optionally the epoch) {of an upcoming solar maximum no later
than right after the start of the given cycle.} This emphasis is also motivated
by the present surge of interest in precisely this topic, prompted by the
unusually long and deep recent solar minimum and by sharply conflicting
forecasts for the maximum of the incipient solar cycle~24.

As we will see, significant doubts arise both from the theoretical and
observational side as to what extent such a prediction is possible at all
(especially before the time of the minimum has become known). Nevertheless, no
matter how shaky their theoretical and empirical backgrounds may be, forecasts
\textit{must} be attempted. Making verifiable or falsifiable predictions is
obviously the core of the scientific method in general; but there is also a more
imperative urge in the case of solar cycle prediction. Being the prime
determinant of space weather, solar activity clearly has enormous technical,
scientific, and financial impact on activities ranging from space exploration to
civil aviation and everyday communication. Political and economic decision
makers expect the solar community to provide them with forecasts on which
feasibility and profitability calculations can be based. Acknowledging this
need, the Space Weather Prediction Center of the US National Weather Service
does present annually or semiannually updated ``official'' predictions of the
upcoming sunspot maximum, emitted by a Solar Cycle Prediction Panel of experts,
starting shortly before the (expected) minimum \citep{SWPC2009}. The unusual
lack of consensus during the early meetings of this panel during the recent
minimum, as well as the concurrent more frequently updated but wildly varying
predictions of a NASA MSFC team \citep{MSFC2009} have put on display the
deficiencies of currently applied prediction techniques; on the other hand, they
also imply that cycle 24 may provide us with crucial new insight into the
physical mechanisms underlying cyclic solar activity.

While a number of indicators of solar activity exist, by far the most commonly
employed is still the smoothed relative sunspot number $R$; the ``Holy Grail''
of sunspot cycle prediction attempts is to get $R$ right for the next maximum.
We therefore start by briefly introducing the sunspot number and inspecting its
known record. Then, in Sections~\ref{sect:precursor},
\ref{sec:Extrapolation-Methods}, and \ref{sec:Model-Based-Predictions}
we discuss the most widely employed methods of cycle
predictions. Section~\ref{sec:Summary-Evaluation} presents a summary
evaluation of the past performance of different forecasting methods
and collects some forecasts for cycle~24 derived by various
approaches. Finally, Section~\ref{sec:Epilogue} concludes the paper.

\subsection{The sunspot number} 
\label{sect:wolfno}

Despite its somewhat arbitrary construction, the series of relative sunspot 
numbers constitutes the longest homogeneous global indicator of solar activity
determined by direct solar observations and carefully controlled methods. For
this reason their use is still predominant in studies of solar activity
variation. As defined originally by \cite{Wolf:relno.def}, the relative sunspot
number is
\begin{equation}
R_W=k(10g+f)  
\end{equation}
where $g$ is the number of sunspot groups (including solitary spots), $f$ is the
total number of all spots visible on the solar disc, while $k$ is a correction
factor depending on a variety of circumstances, such as instrument parameters,
observatory location and details of the counting method. Wolf, who decided to
count each spot only once and not to count the smallest spots, the visibility of
which depended on seeing, used $k=1$. The counting system employed was changed
by Wolf's successors to count even the smallest spots, attributing a higher
weight (i.e.~$f>1$) to spots with a penumbra, depending on their size and umbral
structure. As the new counting naturally resulted in higher values, the
correction factor was set to $k=0.6$ for  subsequent determinations of $R_W$ to
ensure continuity with Wolf's work, even though there was no change in either
the instrument or the observing site. This was followed by several further
changes in the details of the counting method (\citealp{Waldmeier:book}; see
\citealp{Kopecky+:relativeno}, \citealp{Hoyt+Schatten:GSN} and \citealp{Hathaway:LRSP}
for further discussions on the determination of $R_W$). 

In addition to introducing the relative sunspot number, 
\cite{Wolf:cycle.length.activity} also used earlier observational records
available to him to reconstruct its monthly mean values since 1749. In this way,
he reconstructed 11-year sunspot cycles back to that date, introducing their
still universally used numbering. (In a later work he also determined annual
mean values for each calendar year going back to 1700.)

In 1981, the observatory responsible for the official determination of the
sunspot number changed from Z\"urich to the Royal Observatory of Belgium in
Brussels. The website of the SIDC (originally  Sunspot Index Data Center,
recently renamed Solar Influences Data Analysis Center), 
\url{http://sidc.oma.be}, is now the most authoritative source of archive
sunspot number data. But it has to be kept in mind that the sunspot number is
also regularly determined by other institutions: these variants are informally
known as the American sunspot number (collected by AAVSO and available from the
National Geophysical Data Center, \url{http://www.ngdc.noaa.gov/ngdc.html}) and
the Kislovodsk Sunspot Number (available from the web page of the  Pulkovo
Observatory, \url{http://www.gao.spb.ru}). Cycle amplitudes determined by these
other centers may differ by up to 6--7\% from the SIDC values, NOAA numbers
being consistently lower, while Kislovodsk numbers show no such systematic
trend.

These significant disagreements between determinations of $R_W$ by various
observatories and observers are even more pronounced in the case of historical
data, especially prior to the mid-19th century. In particular, the controversial
suggestion that a whole solar cycle may have been missed in the official sunspot
number series at the end of the 18th century is taken by some as glaring
evidence for the unreliability of early observations. Note, however, that
independently of whether the claim for a missing cycle is well founded or not,
there is clear evidence that this controversy is mostly due to the very atypical
behaviour of the Sun itself in the given period of time, rather than to the low
quality and coverage of contemporary observations. These issues will be
discussed further in section \ref{sect:evenodd}.


Given that $R_W$ is subject to large fluctuations on a time scale of days to
months, it has become customary to use annual mean values for the study of
longer term activity changes. To get rid of the arbitrariness of calendar years,
the standard practice is to use 13-month boxcar averages of the monthly averaged
sunspot numbers, wherein the first and last months are given half the weight of
other months:
\begin{equation}
  R=\frac1{24}\left(R_{\mathrm{m},-6}+2\sum_{i=-5}^{i=5} R_{\mathrm{m},i} +R_{\mathrm{m},6}\right)
  \label{eq:Rdef}
\end{equation}
$R_{\mathrm{m},i}$ being the mean monthly value of $R_W$ for $i$th calendar month counted
from the present month. It is this running mean $R$ that we will simply call
``the sunspot number'' throughout this review and what forms the basis of most
discussions of solar cycle variations and their predictions.

In what follows, $\Rmax^{(n)}$ and $\Rmin^{(n)}$ will refer to the maximum and
minimum  value of $R$ in cycle $n$ (the minimum being the one that starts the
cycle). Similarly, $\tmax^{(n)}$ and $\tmin^{(n)}$ will denote the epochs when
$R$ takes these extrema.

\subsubsection{Alternating series and nonlinear transforms} 

Instead of the ``raw'' sunspot number series $R(t)$ many researchers prefer to
base their studies on some transformed index $R'$. The motivation behind this is
twofold.

(a) The strongly peaked and asymmetrical sunspot cycle profiles strongly deviate
from a sinusoidal profile; also the statistical distribution of sunspot numbers
is strongly at odds with a Gaussian distribution. This can constitute a problem
as many common methods of data analysis rely on the assumption of an
approximately normal distribution of errors or nearly sinusoidal profiles of
spectral components. So transformations of $R$ (and, optionally, $t$) that
reduce these deviations can obviously be helpful during the analysis. In this
vein, e.g., Max Waldmeier often based his studies of the solar cycle on the use
of logarithmic sunspot numbers $R'=\log R$; many other researchers use
$R'=R^\alpha$ with $0.5\leq\alpha<1$, the most common value being $\alpha=0.5$.

(b) As the sunspot number is a rather arbitrary construct, there may be an
underlying more physical parameter related to it in some nonlinear fashion, such
as the toroidal magnetic field strength $B$, or the magnetic energy,
proportional to $B^2$. It should be emphasized that, contrary to some claims,
our current understanding of the solar dynamo does \textit{not} make it possible to
guess what the underlying parameter is, with any reasonable degree of certainty.
In particular, the often used assumption that it is the magnetic energy, lacks
any sound foundation. If anything, on the basis of our current best
understanding of flux emergence we might expect that the amount of toroidal flux
emerging from the tachocline should be $\int |B-B_0|\,dA$ where $B_0$ is some
minimal threshold field strength for Parker instability and the surface integral
goes across a latitudinal cross section of the tachocline
\citep[cf.][]{Ruzmaikin:biasing}. As, however, the lifetime of any given sunspot
group is finite and proportional to its size
\citep{Petrovay+vDG:decay1, Henwood+}, instantaneous values of $R$ or
the total sunspot area should also depend on details of the
probability distribution function of $B$ in the tachocline. This just
serves to illustrate the difficulty of identifying a single physical
governing parameter behind $R$.

One transformation that may still be well motivated from the physical point of
view is to attribute an alternating sign to even and odd Schwabe cycles: this
results in the  the \textit{alternating sunspot number series} $R_\pm$. The idea is
based on Hale's well known polarity rules, implying that the period of the solar
cycle is actually 22 years rather than 11 years, the polarity of magnetic fields
changing sign from one 11-year Schwabe cycle to the next. In this
representation, first suggested by \cite{Bracewell}, usually odd cycles are
attributed a negative sign. This leads to slight jumps at the minima of the
Schwabe cycle, as a consequence of the fact that for a 1--2 year period around
the minimum, spots belonging to both cycles are present, so the value of $R$
never reaches zero; in certain applications, further twists are introduced into
the transformation to avoid this phenomenon. 

After first introducing the alternating series, in a later work
\cite{Bracewell:trafo} demonstrated that introducing an underlying ``physical''
variable $R_B$ such that
\begin{equation}
  R_\pm = 100\left( R_B/83\right)^{3/2}
  \label{eq:Bracewell}
\end{equation}
(i.e.~$\alpha=2/3$ in the power law mentioned in in item (a) above)
significantly simplifies the cycle profile. Indeed, upon introducing a
``rectified'' phase variable\epubtkFootnote{The more precise condition defining $\phi$
is that  $\phi=\pm\pi/2$ at each maximum and $\phi$ is quadratically related to
the time since the last minimum.} $\phi$ in each cycle to compensate for the
asymmetry of the cycle profile, $R_B$ is a nearly sinusoidal function of $\phi$.
The empirically found 3/2 law is interpreted as the relation between the
time-integrated area of a typical sunspot group vs. its peak area (or peak $R_W$
value), i.e.~ the steeper than linear growth of $R$ with the underlying physical
parameter $R_B$ would be due to to the larger sunspot groups being observed
longer, and therefore giving a disproportionately larger contribution to the
annual mean sunspot numbers. If this interpretation is correct, as  suggested by
Bracewell's analysis, then $R_B$ should be considered proportional to the total
toroidal magnetic flux emerging into the photosphere in a given interval. (But
the possibility must be kept in mind that that the same toroidal flux bundle may
emerge repeatedly or at different heliographic longitudes, giving rise to
several active regions.)

\subsection{Other indicators of solar activity}

Reconstructions of $R$ prior to the early 19th century are increasingly
uncertain. In order to tackle problems related to sporadic and often unreliable
observations, \cite{Hoyt+Schatten:GSN} introduced the \textit{Group Sunspot Number}
(GSN) as an alternative indicator of solar activity. In contrast to $R_W$, the
GSN only relies on counts of sunspot groups as a more robust indicator,
disregarding the number of spots in each group. Furthermore, while $R_W$ is
determined for any given day from a single observer's measurements (a hierarchy 
of secondary observers is defined for the case if data from the primary observer
were unavailable), the GSN uses a weighted average of all observations
available for a given day. The GSN series has been reproduced for the whole
period 1611--1998 (Figure~\ref{fig:GSNrecord}) and it is generally agreed that for the
period 1611--1818 it is a more reliable reconstruction of solar activity than
the relative sunspot number. Yet there have been relatively few attempts to date
to use this data series for solar cycle prediction. One factor in this could be
the lack of regular updates of the GSN series, i.e. the unavailability of
precise GSN values for the past decade.

\epubtkImage{}{%
  \begin{figure}[htbp]
\centerline{\includegraphics[width=\textwidth]{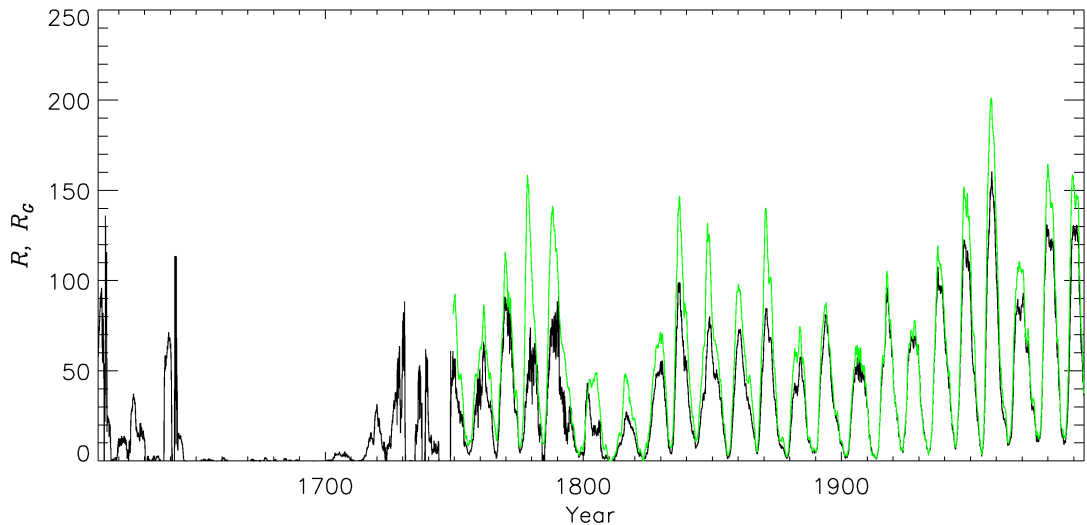}}
\caption{13-month sliding averages of the monthly average relative sunspot
numbers $R$ (green) and group sunspot numbers $R_G$ (black)
for the period 1611--1998.}
\label{fig:GSNrecord}
\end{figure}}


Instead of the sunspot number, the total area of all spots observed on the solar
disk might seem to be a less arbitrary measure of solar activity. However, these
data have been available since 1874 only, covering a much shorter period of time
than the sunspot number data. In addition, the determination of sunspot areas,
especially farther from disk center, is not as trivial as it may seem, resulting
in significant random and systematic errors in the area determinations. Area
measurements performed in two different observatories often show discrepancies
reaching $\sim 30\%$ for smaller spots \citep[cf.\ the figure and discussion in
Appendix~A of][]{Petrovay+:decay2}.

A number of other direct indicators of solar activity have become available from
the 20th century. These include e.g., various plage indices or the 10.7~cm solar
radio flux -- the latter is considered a particularly good and simple to measure
indicator of global activity (cf.\ Figure~\ref{fig:radioflux}). As, however, these
data sets only cover a few solar cycles, their impact on solar cycle prediction
has been minimal.

\epubtkImage{}{%
  \begin{figure}[htbp]
\centerline{\includegraphics[width=\textwidth]{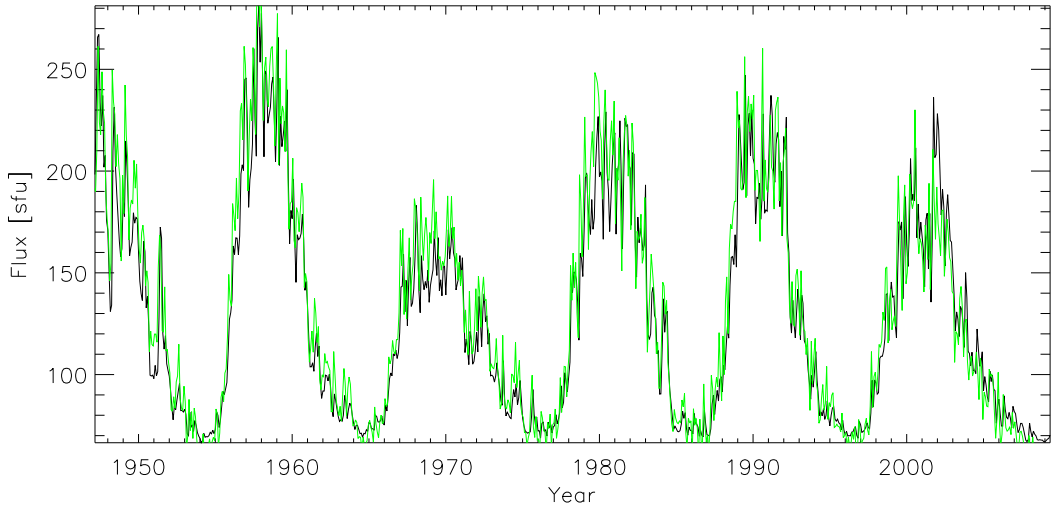}}
\caption{Monthly values of the 10.7~cm radio flux in solar flux units for the
period 1947--2009. The solar flux unit is defined as $10^{-22}\,$W$/$m$^2\,$Hz.
The green curve shows $R_{\mathrm{m}}+60$ where $R_\mathrm{m}$ is the monthly
mean relative sunspot number. (The vertical shift is for better comparison.)
Data are from the NRC Canada (Ottawa/Penticton).}
\label{fig:radioflux}
\end{figure}}

Of more importance are \textit{proxy indicators} such as geomagnetic indices
(the most widely used of which is the $aa$ index), the occurrence frequency of
aurorae or the abundances of cosmogenic radionuclides such as $^{14}$C and
$^{10}$Be. For solar cycle prediction uses such data sets need to have a
sufficiently high temporal resolution to reflect individual 11-year cycles. For
the geomagnetic indices such data have been available since 1868, while an
annual $^{10}$Be series covering 600 years has been published very recently by
\cite{Berggren+:10Be_600yrs}. Attempts have been made to reconstruct the epochs
and even amplitudes of solar maxima during the past two millennia from oriental
naked eye sunspot records and from auroral observations \citep{secular.book,
Nagovitsyn:reconstr}, but these reconstructions are currently subject to too
many uncertainties to serve as a basis for predictions. Isotopic data with lower
temporal resolution are now available for up to 50,000 years in the past; while
such data do not show individual Schwabe cycles, they are still useful for the
study of long term variations in cycle amplitude. Inferring solar activity
parameters from such proxy data is generally not straightforward.

\epubtkImage{}{%
  \begin{figure}[htbp]
\centerline{\includegraphics[width=0.9\textwidth]{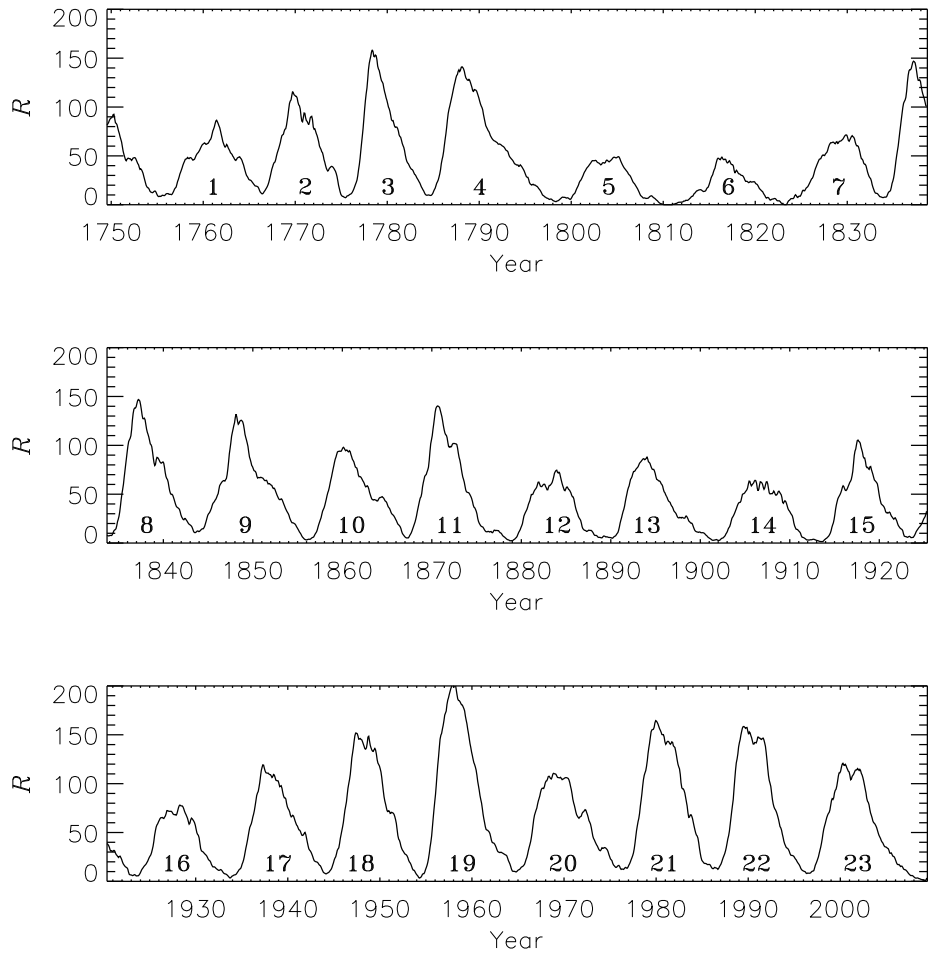}}
\caption{The variation of the monthly smoothed relative sunspot number $R$
during the period 1749--2009, with the conventional numbering of solar
cycles.}
\label{fig:SSNrecord}
\end{figure}}

\subsection{The solar cycle and its variation}
\label{sect:cycle}

The series of $R$ values determined as described in Section~\ref{sect:wolfno} is
plotted in figure~\ref{fig:SSNrecord}. It is evident that the sunspot cycle is
rather irregular. The mean length of a cycle (defined as lasting from minimum to
minimum) is 11.02~years (median 10.7~years), with a standard deviation
of 1.2~years. The mean amplitude is 113 (median 115), with a standard
deviation of 40. It is this wide variation that makes the prediction
of the next cycle maximum such an interesting and vexing issue.

It should be noted that the period covered by the relative sunspot number record
includes an extended interval of atypically strong activity, the so called
Modern Maximum (see below), cycles 17--23. Excluding these cycles from the
averaging, the mean and median values of the cycle amplitude are very close to
100, with a standard deviation of 35. The mean and median cycle length then
become 11.1 and 11.2~years, respectively, with a standard deviation of
1.3~years.

\subsubsection{Secular activity variations}
\label{sect:secular}

Inspecting Figure~\ref{fig:SSNrecord} one can discern an obvious long term
variation. For the study of such long term variations, the series of cycle
parameters is often smoothed on time scales significantly longer than a solar
cycle: this procedure is known as \textit{secular smoothing}. One popular method is
the so-called \textit{Gleissberg filter} or \textit{12221 filter}
\citep{Gleissberg:12221}. For instance, the Gleissberg filtered
amplitude of cycle $n$ is given by
\begin{equation}
\langle \Rmax\rangle_{\mathrm{G}}^{(n)} = \frac 18\left( \Rmax^{(n-2)}
+2\Rmax^{(n-1)} +2\Rmax^{(n)}  +2\Rmax^{(n+1)} +\Rmax^{(n+2)}\right) .
\end{equation}

\epubtkImage{}{%
  \begin{figure}[htbp]
\centerline{\includegraphics[width=\textwidth]{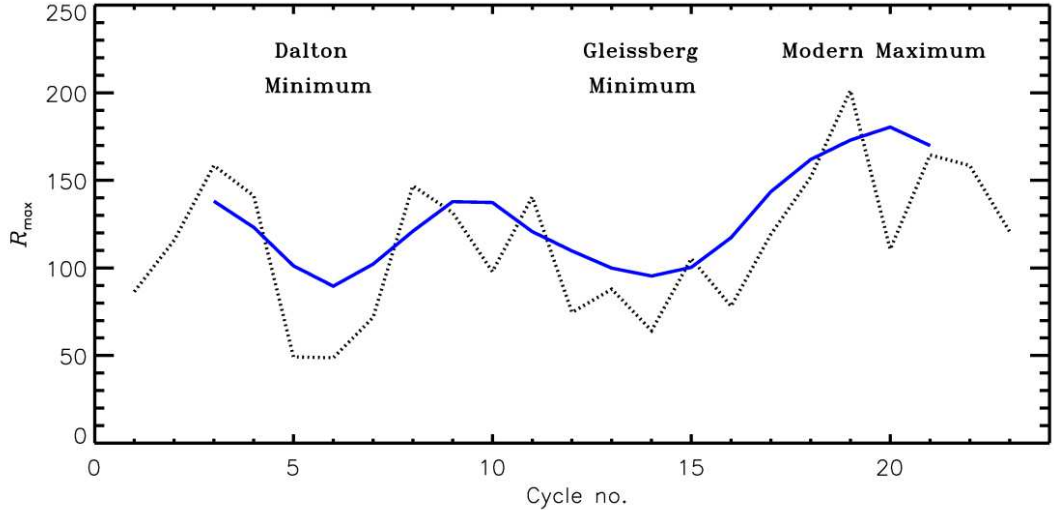}}
\caption{Amplitudes of the sunspot cycles (dotted) and their Gleissberg filtered
values (blue solid), plotted against cycle number. 
}
\label{fig:Gleissbgfilt}
\end{figure}}

The Gleissberg filtered sunspot number series is plotted in Fig.~4. One
long-term trend is an overall secular increase of solar activity, the last six
or seven cycles being unusually strong. (Four of them are markedly stronger than
average and none is weaker than average.) This period of elevated sunspot
activity level from the mid-20th century is known as the ``Modern Maximum''. On
the other hand, cycles 5, 6 and 7 are unusually weak, forming the so-called
``Dalton Minimum''. Finally, the rather long series of moderately weak cycles
12--16 is occasionally referred to as the ``Gleissberg Minimum'' -- but note
that most of these cycles are less than $1\sigma$ below the long-term average.

While the Dalton and Gleissberg minima are but local minima in the ever changing
Gleissberg filtered SSN series, the conspicuous lack of sunspots in the period
1640--1705, known as the Maunder Minimum (Figure~\ref{fig:GSNrecord}) quite
obviously represents a qualitatively different state of solar activity. Such
extended periods of high and low activity are usually referred to as  \textit{grand
maxima} and \textit{grand minima.} Clearly, in comparison with the Maunder
Minimum,  the Dalton Minimum could only be called a ``semi-grand minimum'',
while for the Gleissberg Minimum even that adjective is undeserved.

A number of possibilities have been proposed to explain the phenomenon of grand
minima and maxima, including chaotic behaviour of the nonlinear solar dynamo
\citep{Weiss+:chaotic.dynamo}, stochastic fluctuations in dynamo parameters
\citep{Moss+:stochastic.dynamo1, Moss+:stochastic.dynamo2} or a
bimodal dynamo with stochastically induced alternation between two stationary
states \citep{Petrovay:bimodal}.

The analysis of long-term proxy data, extending over several millennia further
showed that there exist systematic long-term statistical trends and periods such
as the so called  secular and supersecular cycles (see
Section~\ref{sect:spectral}).

\subsubsection{Does the Sun have a long term memory? \label{sect:memory}}

Following customary usage, by ``memory'' we will refer to some physical (or, in
the case of a model, mathematical) mechanism by which the state of a system at a
given time will depend on its previous states. In any system there may be
several different such mechanisms at work simultaneously ---if this is so, again
following common usage we will speak of different ``types'' of memory. A very
mundane example are the RAM and the hard disk in a computer: devices that store
information over very different time scales and the effect of which manifests
itself differently in the functioning of the system.

There is no question that the solar dynamo (i.e. the mechanism that gives rise
to the sunspot number series) does possess a memory that extends at least over
the course of a single solar cycle. Obviously, during the rise phase solar
activity ``remembers'' that it should keep growing, while in the decay phase it
keeps decaying, even though exactly the same range of $R$ values are observed in
both phases. Furthermore, profiles of individual sunspot cycles may, in a first
approximation, be considered a one-parameter ensemble
\citep{Hathaway:periodampl}.  This obvious effect will be referred to here as
{\it intracycle memory.} 

As we will see, correlations between activity parameters in different cycles are
generally much weaker than those within one cycle, which strongly suggests that
the intracycle memory mechanism is different from longer term memory effects, if
such are present at all. Referring back to our analogy, the intracycle memory
may work like computer RAM, periodically erased at every reboot (i.e. at the
start of a new cycle).

The interesting question is whether, in addition to the intracycle memory
effect, any other type of memory is present in the solar dynamo or not. To what
extent is the amplitude of a sunspot cycle determined by previous cycles? Are
subsequent cycles essentially independent, randomly drawn from some stochastic
distribution of cycle amplitudes around the long term average? Or, in the
alternative case, for how many previous cycles do we need to consider solar
activity for successful forecasts? 

The existence of long lasting grand minima and maxima suggests that the sunspot
number record must have a {\it long-term memory} extending over several
consecutive cycles. Indeed, elementary combinatorical calculations show that the
occurrence of phenomena like the Dalton minimum (3 of the 4 lowest maxima
occurring in a row) or the Modern maximum (4 of the 5 highest maxima occurring
within a series of 5 cycles) in a random series of 24 recorded solar maxima has
a rather low probability (5\,\% and 3\,\%, respectively). This conclusion is
corroborated by the analysis of long-term proxy data, extending over several
millennia, which showed that the occurrence of grand minima and grand maxima is
more common than what would follow from Gaussian statistics
\citep{Usoskin+Solanki}.

It could be objected that for sustained grand minima or maxima a memory
extending only from one cycle to the next would suffice. In contrast to 
long-term (multidecadal or longer) memory, this would constitute another kind of
short-term ($\la 10$ years) memory: a cycle-to-cycle or {\it intercycle} memory
effect. In our computer analogy, think of system files or memory cache
written on the hard disk, often with the explicit goal of recalling the system
status (e.g. desktop arrangement) after the next reboot. While these files
survive the reboot, they are subject to erasing and rewriting in every session,
so they have a much more temporary character than the generic data files stored
on the disk.

The intercycle memory explanation of persistent secular activity minima and
maxima, however, would imply a good correlation between the amplitudes of
subsequent cycles, which is not the case (cf.\ Sect.\ \ref{sect:minimax} below).
With the known poor cycle-to-cycle correlation, strong deviations from the
long-term mean would be expected to be damped on time scales short compared to
e.g.\ the length of the Maunder minimum. This suggests that the persistent
states of low or high activity are due to truly long term memory effects
extending over several cycles.

Further evidence for a long-term memory in solar activity comes from the
persistence analysis of activity indicators. The parameter determined in such
studies is the Hurst exponent $0<H<1$. Essentially, $H$ is the steepness of the
growth of the total range $\cal R$ of measured values plotted against the number
$n$ of data in a time series, on a logarithmic plot: ${\cal R}\propto n^H$.  For
a Markovian random process with no memory $H=0.5$. Processes with $H>0.5$ are
persistent (they tend to stay in a stronger-than-average or weaker-than-average
state longer), while those with $H<0.5$ are anti-persistent (their fluctuations
will change sign more often). 

Hurst exponents for solar activity indices have been derived using rescaled
range analysis by many authors \citep{Mandelbrot+:Hurst,
  Ruzmaikin:Hurst, Komm:DopplerHurst, Oliver+Ballester:Hurst,
  Kilcik+}. All studies coherently yield a value $H=0.85$--$0.88$ for
time scales exceeding a year or so, and somewhat lower values ($H\sim
0.75$) on shorter time scales. Some doubts regarding the significance
of this result for a finite series have been raised by
\cite{Oliver+Ballester:noHurst}; however, \cite{Qian+Rasheed} have
shown using Monte-Carlo experiments that for time series of a length
comparable to the sunspot record, $H$ values exceeding 0.7 are
statistically significant. 

A complementary method, essentially equivalent to rescaled range analysis is
detrended fluctuation analysis. Its application to solar data
\citep{Ogurtsov:Hurst} has yielded results in accordance with the $H$
values quoted above.

The overwhelming evidence for the persistent character of solar activity and for
the intermittent appearance of secular cyclicities, however, is not much help
when it comes to cycle-to-cycle prediction. It is certainly reassuring to know
that forecasting is not a completely idle enterprise (which would be the case
for a purely Markovian process), and the long-term persistence and trends may
make our predictions statistically somewhat different from just the long-term
average.  There are, however, large decadal scale fluctuations superposed on the
long term trends, so the associated errors will still be so large as to make the
forecast of little use for individual cycles.

\epubtkImage{}{%
  \begin{figure}[htbp]
\centerline{\includegraphics[width=\textwidth]{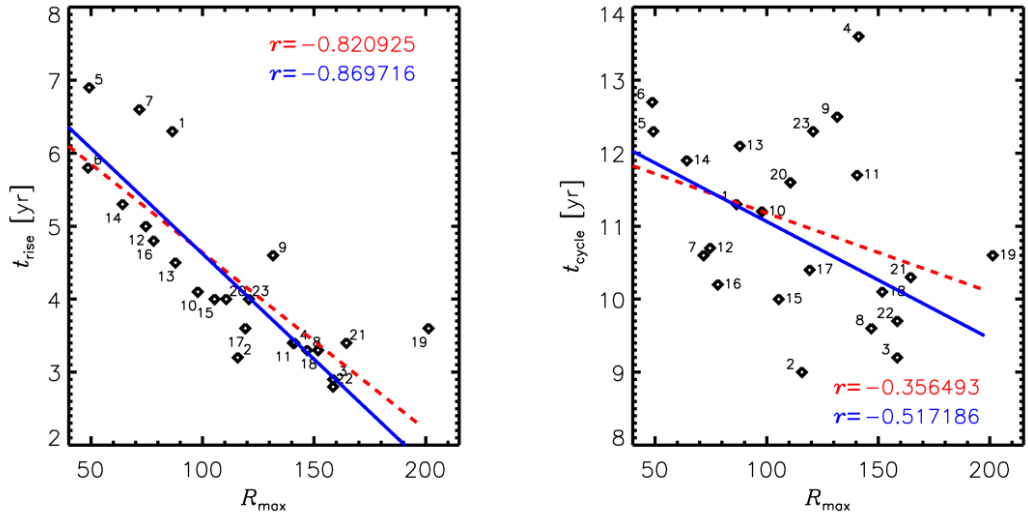}}
\caption{Monthly smoothed sunspot number $R$ at cycle maximum plotted against the rise
time to maximum (left) and against cycle length (right). Cycles are labeled with
their numbers. In the plots the red dashed lines are linear regressions
to all the data, while the blue solid lines are fits to all data except
outliers. Cycle~19 is considered an outlier on both plots, cycle~4 on the right
hand plot only.  The corresponding correlation coefficients are shown.}
\label{fig:waldmeier}
\end{figure}}

\subsubsection{Waldmeier effect and amplitude--frequency correlation}
\label{sect:Waldmeier}

\begin{nquote}
{\sl  Greater activity on the Sun goes with shorter periods, and less with
longer periods. I believe this law to be one of the most important relations
among the Solar actions yet discovered.}\\
\strut\hfill{\citep{Wolf:cycle.length.activity}}
\end{nquote}

It is apparent from Figure~\ref{fig:SSNrecord} that the profile of sunspot cycles
is asymmetrical, the rise being steeper than the decay. Solar activity maxima
occur 3 to 4 years after the minimum, while it takes another 7--8 years to reach
the next minimum. It can also be noticed that the degree of this asymmetry
correlates with the amplitude of the cycle: to be more specific, the length of
the rise phase anticorrelates with the maximal value of $R$
(Figure~\ref{fig:waldmeier}), while the length of the decay phase shows weak or no
such correlation. 

Historically, the relation was first formulated by \cite{Waldmeier:effect} as an
inverse correlation between the rise \textit{time} and the cycle amplitude;
however, as shown by \cite{Tritakis:Waldmeier}, the total rise time is a weak
(inverse logarithmic) function of the rise rate, so this representation makes
the correlation appear less robust. (Indeed, when formulated with the rise time
it is not even present in some activity indicators, such as sunspot
areas -- cf.\ \cite{Dikpati+:Waldmaier}.)  As pointed out by \cite{Cameron+:Waldmeier}, the
weak link between rise time and slope is due to the fact that in steeper rising
cycles the minimum will occur earlier, thus partially compensating for the
shortening due to a higher rise rate. The effect is indeed more clearly seen
when the rate of the rise is used instead of the rise time
\citep{Lantos, Cameron+:Waldmeier}. The observed correlation between rise rate and
maximum cycle amplitude is approximately linear, good (correlation coefficient
$r\sim 0.85$), and quite robust, being present in various activity indices.

Nevertheless, when coupled with the nearly nonexistent correlation between the
decay time and the cycle amplitude, even the weaker link between the rise time
and the maximum amplitude is sufficient to forge a weak inverse correlation
between the total cycle length and the cycle amplitude
(Figure~\ref{fig:waldmeier}). This inverse relationship was first noticed by
\cite{Wolf:cycle.length.activity}.

A stronger inverse correlation was found between the cycle amplitude and the
length of the {\it previous} cycle by \cite{Hathaway:periodampl}. This
correlation is also readily explained as a consequence of the Waldmeier effect,
as demonstrated in a simple model by \cite{Cameron+:prediction}. Note that in a
more detailed study \cite{Solanki+:cycle.length} find that the correlation
coefficient of this relationship has steadily decreased during the course of the
historical sunspot number record, while the correlation between cycle amplitude
and the length of the {\it third} preceding cycle has steadily increased. The
physical significance (if any) of this latter result is unclear. 

In what follows, the relationships found by \cite{Wolf:cycle.length.activity},
\cite{Hathaway:periodampl}, and \cite{Solanki+:cycle.length}, discussed above,
will be referred to as ``$\Rmax$--$t_{\mathrm{cycle},n}$ correlations'' with
$n=0$, $-1$ or $-3$, respectively.

Modern time series analysis methods offer several ways to define an
instantaneous frequency $f$ in a quasiperiodic series. One simple approach was
discussed in the context of Bracewell's transform, Equation~(\ref{eq:Bracewell}),
above. \cite{Mininni+:vanderpol} discuss several more sophisticated methods to
do this, concluding that G\'abor's analytic signal approach yields the best
performance. This technique was first applied to the sunspot record by
\cite{Palus+Novotna}, who found a significant long term correlation between the
smoothed instantaneous frequency and amplitude of the signal. On time scales
shorter than the cycle length, however, the frequency--amplitude correlation has
not been convincingly proven, and the fact that the correlation coefficient is
close to the one reported in the right hand panel of Figure~\ref{fig:waldmeier}
indicates that all that the fashionable gadgetry of nonlinear dynamics could
achieve was to recover the effect already known to Wolf. It is clear from this
that the ``frequency--amplitude correlation'' is but a secondary consequence of
the Waldmeier effect.

On the left hand panel of Figure~\ref{fig:waldmeier}, within the band of
correlation the points seem to be sitting neatly on two parallel strings. Any
number of faint hearted researchers would dismiss this as a coincidence or as
another manifestation of the ``Martian canal effect''. But
\cite{Kuklin:Waldmeier} boldly speculated that the phenomenon may be real. Fair
enough, cycles 22 and 23 dutifully took their place on the lower string even
after the publication of Kuklin's work. This speculation was supported with
further evidence by \cite{Nagovitsyn:reconstr} who offered a physical
explanation in terms of the amplitude--frequency diagram of a forced nonlinear
oscillator (cf.\ Section~\ref{sect:oscillator}).

Indeed, an anticorrelation between cycle length and amplitude is characteristic
of a class of stochastically forced nonlinear oscillators and it may also be
reproduced by introducing a stochastic forcing in dynamo models
\citep{Stix:Waldmeier, Ossendrijver+:stoch.dynamo, Charbonneau+Dikpati}. In some
such models the characteristic asymmetric profile of the cycle is also well
reproduced \citep{Mininni+:vanderpol, Mininni+:vanderpol2}. The predicted
amplitude--frequency relation has the form
\begin{equation}
  \log\Rmax^{(n)} = C_1+ C_2f \,.
  \label{eq:stochRf}
\end{equation}

Nonlinear dynamo models including some form of $\alpha$-quenching also have the
potential to reproduce the effects described by Wolf and Waldmeier without
recourse to stochastic driving. In a dynamo with a Kleeorin--Ruzmaikin type
feedback on $\alpha$, \cite{Kitiashvili+:nonlin.dynamo} are able to
qualitatively reproduce the Waldmeier effect. Assuming that the sunspot number
is related to the toroidal field strength according to the Bracewell transform,
Equation~(\ref{eq:Bracewell}), they find a strong link between rise time and
amplitude, while the correlations with fall time and cycle length are much
weaker, just as the observations suggest. They also find that the form of the
growth time--amplitude relationship differs in the regular (multiperiodic) and
chaotic regimes. In the regular regime the plotted relationship
suggests
\begin{equation}
\Rmax^{(n)} = C_1-C_2\left(\tmax^{(n)}-\tmin^{(n)}\right) ,
\end{equation}
while in the chaotic case  
\begin{equation}
\Rmax^{(n)} \propto {\left[1/\left(\tmax^{(n)}-\tmin^{(n)}\right)\right]} .
\end{equation}

Note that based on the actual sunspot number series Waldmeier originally
proposed
\begin{equation}
\log\Rmax^{(n)} =C_1-C_2\left(\tmax^{(n)}-\tmin^{(n)}\right) , 
\end{equation}
%
while according to \cite{Dmitrieva:Waldmeierdynamo} the relation takes the
form
\begin{equation}
\log\Rmax^{(n)} \propto {\left[1/\left(\tmax^{(n)}-\tmin^{(n)}\right)\right]} . 
\end{equation}

At first glance, these logarithmic empirical relationships seem to be more
compatible with the relation (\ref{eq:stochRf}) predicted by the stochastic
models. These, on the other hand, do not actually reproduce the Waldmeier
effect, just a general asymmetric profile and an amplitude--frequency
correlation. At the same time, inspection of the the left hand panel in
Figure~\ref{fig:waldmeier} shows that the data is actually not incompatible with a
linear or inverse rise time--amplitude relation, especially if the anomalous
cycle~19 is ignored as an outlier. (Indeed, a logarithmic representation is
found not to improve the correlation coefficient -- its only advantage is that
cycle~19 ceases to be an outlier.) All this indicates that nonlinear dynamo
models may have the potential to provide a satisfactory quantitative explanation
of the Waldmeier effect, but more extensive comparisons will need to be done,
using various models and various representations of the relation.


\newpage

\section{Precursor Methods}
\label{sect:precursor}

\begin{nquote}
{\sl Jeder Fleckenzyklus mu{\ss} als ein abgeschlossenes Ganzes, als ein
Ph{\"a}nomen f{\"u}r sich, aufgefa{\ss}t werden, und es reiht sich einfach
Zyklus an Zyklus.}\\ 
\strut\hfill{\citep{Gleissberg:book}}
\end{nquote}

In the most general sense, precursor methods rely on the value of some measure
of solar activity or magnetism at a specified time to predict the amplitude of
the following solar maximum. The precursor may be any proxy of solar activity or
other indicator of solar and interplanetary magnetism. Specifically, the
precursor may also be the value of the sunspot number at a given time.

In principle, precursors might also herald the activity level at other phases of
the sunspot cycle, in particular the minimum. Yet the fact that practically all
the good precursors found need to be evaluated at around the time of the minimum
and refer to the next maximum is not simply due to the obvious greater interest
in predicting maxima than predicting minima. Correlations between minimum
parameters and previous values of solar indices have been looked for, but the
results were overwhelmingly negative \citep[e.g.][]{Tlatov:polar.precursors}. This
indicates that the sunspot number series is \textit{not} homogeneous and Rudolf
Wolf's instinctive choice to start new cycles with the minimum rather than the
maximum in his numbering system is not arbitrary -- for which even more obvious
evidence is provided by the butterfly diagram. Each numbered solar cycle is a
consistent unit in itself, while solar activity seems to consist of a series of
much less tightly intercorrelated individual cycles, as suggested by Wolfgang
Gleissberg in the motto of this section.

In Section \ref{sect:memory} we have seen that there is significant evidence for
a {\it long-term memory} underlying solar activity. In addition to the evidence
reviewed there, systematic long-term statistical trends and periods of solar
activity, such as the secular and supersecular cycles (to be discussed in
Section~\ref{sect:spectral}) also attest to a secular mechanism underlying solar
activity variations and ensuring some degree of long-term coherence in activity
indicators. However, as we noted, this long-term memory is of limited importance
for cycle prediction due to the large, apparently haphazard decadal variations
superimposed on it. What the precursor methods promise is just to find a system
in those haphazard decadal variations ---which clearly implies a different type
of memory. As we already mentioned in Section \ref{sect:memory}, there is
obvious evidence for an {\it intracycle memory} operating \textit{within} a
single cycle, so that forecasting of activity in an ongoing cycle is currently a
much more successful enterprise than cycle-to-cycle forecasting. As we will see,
this intracycle memory is one candidate mechanism upon which precursor
techniques may be founded, via the Waldmeier effect.


The controversial issue is whether, in addition to the intracycle memory, there
is also an {\it intercycle memory} at work, i.e. whether behind the apparent
stochasticity of the cycle-to-cycle variations there is some predictable
pattern, whether some imprint of these variations is somehow inherited from one
cycle to the next, or individual cycles are essentially independent. The latter
is known as the ``outburst hypothesis'': consecutive cycles would then represent
a series of ``outbursts'' of activity with stochastically fluctuating amplitudes
(\citealp{Halm, Waldmeier:effect, Vitinsky:book2}; see also
\citealp{deMeyer:impulse} who calls this ``impulse model''). Note that
cycle-to-cycle predictions in the strict temporal sense may be possible even in
the outburst case, as solar cycles are known to overlap. Active regions
belonging to the old and new cycles may coexist for up to three years or so
around sunspot minima; and high latitude ephemeral active regions oriented
according to the next cycle appear as early as 2--3 years after the maximum
(\citealp{Tlatov+:ER} -- the so-called extended solar cycle).

In any case, it is undeniable that for cycle-to-cycle predictions, which are our
main concern here, the precursor approach seems to have been the relatively most
successful, so its inherent basic assumption must contain an element of truth --
whether its predictive skill is due to a ``real'' cycle-to-cycle memory
(intercycle memory) or just to the overlap effect (intracycle memory).

The two precursor types that have received most attention are polar field
precursors and geomagnetic precursors. A link between these two categories is
forged by a third group, characterizing the interplanetary magnetic field
strength or ``open flux''. But before considering these approaches, we start by
discussing the most obvious precursor type: the level of solar activity at some
epoch before the next maximum.



\epubtkImage{}{%
  \begin{figure}[htbp]
    \centerline{\includegraphics[width=\textwidth]{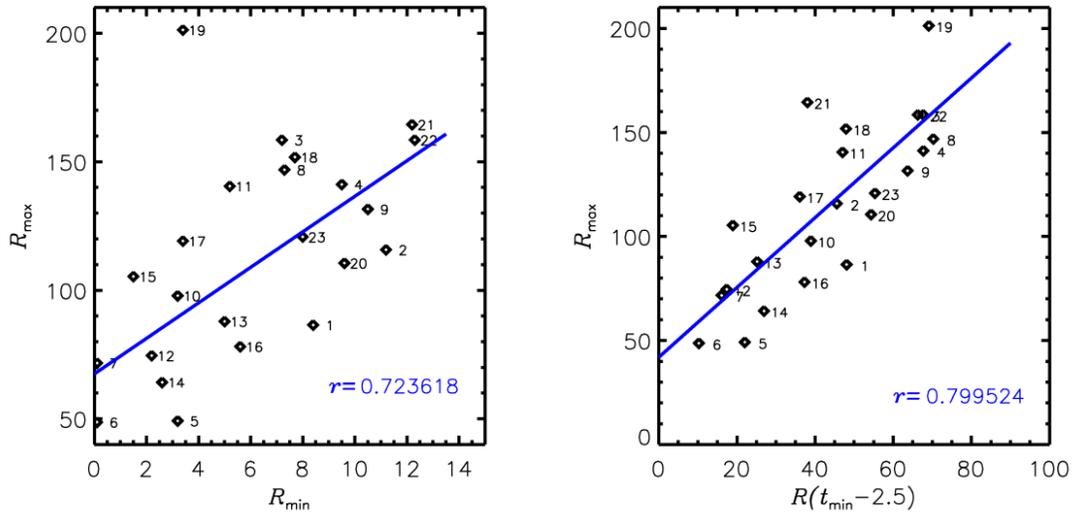}}
    \caption{Monthly smoothed sunspot number $R$ at cycle maximum
      plotted against the values of $R$ at the previous minimum (left)
      and 2.5 years before the minimum (right). Cycles are labeled
      with their numbers. The blue solid line is a linear regression
      to the data; corresponding correlation coefficients are
      shown. In the left hand panel, cycle~19 was considered an
      outlier.}
    \label{fig:minimax}
\end{figure}}

\subsection{Cycle parameters as precursors and the Waldmeier effect}
\label{sect:minimax}

The simplest weather forecast method is saying that ``tomorrow the weather will
be just like today'' (works in about 2/3 of the cases). Similarly, a simple
approach of sunspot cycle prediction is correlating the amplitudes of
consecutive cycles. There is indeed a marginal correlation, but the correlation
coefficient is quite low (0.35). The existence of the correlation is related to
secular variations in solar activity, while its weakness is due to the
significant cycle-to-cycle variations.

A significantly better correlation exists between the \textit{minimum} activity
level and the the amplitude of the next maximum (Figure~\ref{fig:minimax}). The
relation is linear \citep{Brown:minimax}, with a correlation coefficient of
0.72 (if the anomalous cycle 19 is ignored -- \citealp{Brajsa+:gleissbg}; see also
\citealp{Pishkalo}). The best fit is
\begin{equation}
\Rmax=67.5+6.91 \Rmin
\label{eq:minimax}
\end{equation}
Using the observed value 1.7 for the SSN in the recent minimum, the next maximum
is predicted by this ``minimax'' method to reach values around 80, with a
$1\sigma$ error of about $\pm 25$. 

\cite{Cameron+:prediction} point out that the activity level three years before
the minimum is an even better predictor of the next maximum. Indeed, playing
with the value of time shift we find that the best correlation coefficient
corresponds to a time shift of 2.5 years, as shown in.the right hand panel of
Figure~\ref{fig:minimax} (but this may depend on the particular time period
considered, so we will refer to this method in Table~1 as ``minimax3'' for
brevity). The linear regression is
\begin{equation}
\Rmax=41.9+1.68 R(\tmin-2.5) .
\label{eq:minimax3}
\end{equation}
For cycle~24 the value of the predictor is 16.3, so this indicates an amplitude
of 69, suggesting that the upcoming cycle may be comparable in strength to
those during the Gleissberg minimum at the turn of the 19th and 20th centuries.

As the epoch of the minimum of $R$ cannot be known with certainty until about a
year \textit{after} the minimum, the practical use of these methods is rather
limited: a prediction will only become available 2--3 years before the maximum,
and even then with the rather low reliability reflected in the correlation
coefficients quoted above. In addition, as convincingly demonstrated by
\cite{Cameron+:prediction} in a Monte Carlo simulation, these methods do not
constitute real cycle-to-cycle prediction in the physical sense: instead, they
are due to a combination of the overlap of solar cycles with the Waldmeier
effect. As stronger cycles are characterized by a steeper rise phase, the
minimum before such cycles will take place earlier, when the activity from the
previous cycle has not yet reached very low levels.

The same overlap readily explains the $\Rmax$--$t_{\mathrm{cycle},n}$
correlations discussed in Section \ref{sect:Waldmeier}. These relationships may
also be used for solar cycle prediction purposes \citep[e.g.][]{Kane:cyclength}
but they lack robustness. For cycle 24 the $\Rmax$--$t_{\mathrm{cycle},-1}$
correlation, as formulated by \cite{Hathaway:LRSP} predicts $\Rmax=80$ while the
methods used by \cite{Solanki+:cycle.length} yield values ranging from 86 to
about 110, depending on the relative weights of $t_{\mathrm{cycle},-1}$ and
$t_{\mathrm{cycle},-3}$. The forecast is not only sensitive to the value of $n$
used but also to the data set (relative or group sunspot numbers)
\citep{Vaquero+:cyclength}.

\epubtkImage{}{%
  \begin{figure}[htbp]
\centerline{\includegraphics[width=\textwidth]{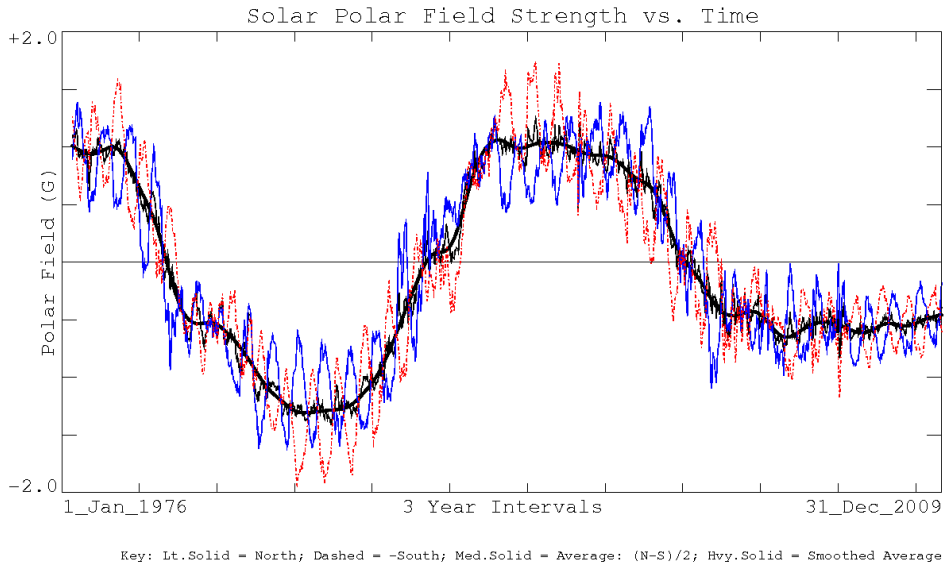}}
\caption{Magnetic field strength in the Sun's polar regions as a function of
time. Blue solid: North; red dashed: $(-1)\cdot$South; thin black solid:
average; heavy black solid: smoothed average). Strong annual modulations in the
hemispheric data are due to the tilt of the solar equator to the Ecliptic. Data
and figure: Wilcox Solar Observatory -- see
\url{http://wso.stanford.edu/gifs/Polar.gif} for updated version}
\label{fig:polarfield}
\end{figure}}

\subsection{Polar precursors}
\label{sect:polar}

Direct measurements of the magnetic field in the polar areas of the Sun have
been available from Wilcox Observatory since 1976
\citep{Svalgaard+:polarfield, Hoeksema:SSR}. Even before a significant
amount of data had been available for statistical analysis, solely on the basis
of the Babcock--Leighton scenario of the origin of the solar cycle,
\cite{Schatten+:polar.prec} suggested that the polar field measurements may be
used to predict the amplitude of the next solar cycle. Data collected in the four
subsequent solar cycles have indeed confirmed this suggestion. As it was
originally motivated by theoretical considerations, this \textit{polar field
precursor method} might also be a considered a model-based prediction technique.
As, however, no particular detailed mathematical model is underlying the method,
numerical predictions must still be based on empirical correlations --
hence our categorization of this technique as a precursor method.

The shortness of the available direct measurement series represents a difficulty
when it comes to finding empirical correlations to solar activity. This problem
can to some extent be circumvented by the use of proxy data. For instance,
\cite{Obridko+Shelting} use H$_\alpha$ synoptic maps to reconstruct the polar
field strength at the source surface back to 1915. Spherical harmonic expansions
of global photospheric magnetic measurements can also be used to deduce the
field strength near the poles.  The use of such proxy techniques permits a
forecast with a sufficiently restricted error bar to be made, despite the
shortness of the direct polar field data set.

The polar fields reach their maximal amplitude near minima of the sunspot cycle.
In its most commonly used form, the polar field precursor method employs the
value of the polar magnetic field strength (typically, the absolute value of the
mean field strength poleward of 55$^\circ$ latitudes, averaged for the two
hemispheres) at the time of sunspot minimum. It is indeed remarkable that
despite the very limited available experience, forecasts using the polar field
method have proven to be consistently in the right range for cycles 21, 22 and
23 \citep{Schatten+:pred22, Schatten+:pred23}. 

By virtue of the definition~(\ref{eq:Rdef}), the time of the minimum of $R$
cannot be known earlier than 6 months \textit{after} the minimum  -- indeed, to
make sure that the perceived minimum was more than just a local dip in $R$, at
least a year or so needs to elapse. This would suggest that the predictive value
of polar field measurements is limited, the prediction becoming available 2--3
years before the upcoming maximum only. 

To remedy this situation, \cite{Schatten+:SoDA} introduced a new activity index,
the ``Solar Dynamo Amplitude'' (SoDA) index, combining the polar field strength
with a traditional activity indicator (the 10.7~cm radio flux $F10.7$). Around
minimum, SoDA is basically proportional to the polar precursor and its value
yields the prediction for $F10.7$ at the next maximum; however, it was
constructed so that its 11-year modulation is minimized, so theoretically it
should be rather stable, making predictions possible well before the minimum.
That's the theory, anyway -- in reality, SoDA based forecasts made more than
2--3 years before the minimum usually proved unreliable. It is then questionable
to what extent SoDA improves the prediction skill of the polar precursor, to
which it is more or less equivalent in those late phases of the solar cycle when
forecasts start to become reliable.

Fortunately, however, the maxima of the polar field curves are often rather flat
(cf.\ Figure~\ref{fig:polarfield}), so approximate forecasts are feasible several
years before the actual minimum. Using the current, rather flat and low maximum
in polar field strength, \cite{Svalgaard+:prediction24} have been able to
predict a relatively weak cycle 24 (peak $R$ value $75\pm 8$) as early as 4
years before the sunspot minimum took place in December 2008! Such an early
prediction is not always possible: early polar field predictions of cycles 22
and 23 had to be corrected later and only forecasts made shortly before the
actual minimum did finally converge. Nevertheless, even the moderate success
rate of such early predictions seems to indicate that the suggested physical
link between the precursor and the cycle amplitude is real.

In addition to their above mentioned use in reconstructing the polar field
strength, various proxies or alternative indices of the global solar
magnetic field during the activity minimum may also be used directly as
activity cycle precursors. H$_\alpha$ synoptic charts are  now
available from various observatories from as early as 1870. As H$_\alpha$
filaments lie on the magnetic neutral lines, these maps can be used to
reconstruct the overall topology, if not the detailed map, of the large-scale
solar magnetic field. \cite{Tlatov:polar.precursors} has shown that several
indices of the polar magnetic field during the activity minimum, determined from
these charts, correlate well with the amplitude of the incipient cycle.

High resolution Hinode observations have now demonstrated that the polar
magnetic field has a strongly intermittent structure, being concentrated in
intense unipolar tubes that coincide with polar faculae
\citep{Tsuneta+:polar.landscape}. The number of polar faculae should then also
be a plausible proxy of the polar magnetic field strength and a good precursor
of the incipient solar cycle around the minimum. This conclusion was indeed
confirmed by \cite{Li+:polarfac}, and, more recently, by
\cite{Tlatov:polar.precursors}.

These methods offer a prediction over a time span of 3--4 years, comparable to
the rise time of the next cycle.  A significantly earlier prediction possibility
was, however, suggested by \cite{Makarov:polfacpred89} and
\cite{Makarov+:polfacpred96} based on the number of polar faculae observed at
Kislovodsk, which was found to predict the next sunspot cycle with a time lag of
5--6 years; even short term annual variations or ``surges'' of sunspot activity
were claimed to be discernible in the polar facular record. This surprising
result may be partly due to the fact that Makarov {et~al.} considered all
faculae poleward of 50 degrees latitude. \textit{Bona fide} polar faculae, seen
on Hinode images to be knots of the unipolar field around the poles, are limited
to higher latitudes, so the wider sample may consist of a mix of such ``real''
polar faculae and small bipolar ephemeral active regions. These latter are known
to obey an extended butterfly diagram, as recently confirmed by
\cite{Tlatov+:ER}: the first bipoles of the new cycle appear at higher latitudes
about 4 years after the activity maximum. It is not impossible that these early
ephemeral active regions may be used for prediction purposes \citep[cf.~also
][]{Badalyan+}; but whether or not the result of \cite{Makarov:polfacpred89}
may be attributed to this is doubtful, as \cite{Li+:polarfac} find that even
using all polar faculae poleward of $50^\circ$ from the Mitaka data base, the
best autocorrelation still results with a time shift of about 4 years only.

Finally, trying to correlate various solar activity parameters,
\cite{Tlatov:polar.precursors} finds this surprising relation:
\begin{equation}
  \Rmax^{(n+1)}=C_1+C_2\Rmax^{(n)} 
  \left(t_\mathrm{rev}^{(n)}-\tmax^{(n)}\right)
  \qquad C_1=83\pm 11 \quad C_2=0.09\pm 0.02
  \label{eq:tlatov}
\end{equation}
where $t_\mathrm{rev}^{(n)}$ is the epoch of the polarity reversal in cycle $n$
(typically, about a year after $t_\mathrm{max}^{(n)}$). The origin of this
curious relationship is unclear. In any case, the good correlation coefficient
($r=0.86$, based on 12 cycles) and the time lag of $\sim 10$ years make this
relationship quite remarkable. For cycle 24 this formula predicts
$R_{\mathrm{max}}^{(24)}=94\pm 14$

\subsection{Geomagnetic and interplanetary precursors}
\label{sect:geomg}

Relations between the cycle related variations of geomagnetic indices and solar
activity were noted long ago. It is, however, important to realize that the
overall correlation between geomagnetic indices and solar activity, even after
13-month smoothing, is generally far from perfect. This is due to the fact that 
the Sun can generate geomagnetic disturbances in two ways:

\begin{descr}
\item{(a)} By material ejections (such as CMEs or flare particles) hitting the
terrestrial magnetosphere. This effect is obviously well correlated with solar
activity, with no time delay, so this contribution to geomagnetic disturbances
peaks near, or a few years after, sunspot maximum. (Note that the occurrence of
the largest flares and CMEs is known to peak some years after the sunspot
maximum -- see Figure~16 in \citealp{Hathaway:LRSP}.)
\item{(b)} By a variation of the strength of the general interplanetary magnetic
field and of solar wind speed. Geomagnetic disturbances may be triggered by the
alternation of the Earth's crossing of interplanetary sector boundaries (slow
solar wind regime) and its crossing of high speed solar wind streams while well
within a sector. The amplitude of such disturbances will clearly be higher for
stronger magnetic fields. The overall strength of the interplanetary magnetic
field, in turn, depends mainly on the total flux present in coronal holes, as
calculated from potential field source surface models of the coronal magnetic
field. At times of low solar activity the dominant contribution to this flux
comes from the two extended polar coronal holes, hence, in a simplistic
formulation this interplanetary contribution may be considered linked to the
polar magnetic fields of the Sun, which in turn is a plausible precursor
candidate as we have seen in the previous subsection. As the polar field
reverses shortly after sunspot maximum, this second contribution often
introduces a characteristic secondary minimum in the cycle variation of
geomagnetic indices, somewhere around the maximum of the curve.
\end{descr}

The component (a) of the geomagnetic variations actually \textit{follows} sunspot
activity with a variable time delay. Thus a geomagnetic precursor based on
features of the cycle dominated by this component has relatively little 
practical utility. This would seem to be the case, e.g., with the forecast method
first proposed by \cite{Ohl66}, who noticed that the minimum amplitudes of the
smoothed geomagnetic $aa$ index are correlated to the amplitude of the next
sunspot cycle \citep[see also][]{Du+:Ohlmethod}.

An indication that the \textit{total} geomagnetic activity, resulting from both
mechanisms does contain useful information on the expected amplitude of the next
solar cycle was given by \cite{Thompson:geomg.prec}, who found that the total
number of disturbed days in the geomagnetic field in cycle $n$ is related to the
sum of the amplitudes of cycles $n$ and $n+1$. \citep[See also][]{Dabas+}.

A method for separating component (b) was proposed by \cite{Feynman} who
correlated the annual $aa$ index with the annual mean sunspot number and found a
linear relationship between $R$ and the \textit{minimal} value of $aa$ for years
with such $R$ values. She interpreted this linear relationship as representing
the component (a) discussed above, while the amount by which $aa$ in a given
year actually exceeds the value predicted by the linear relation would be the
contribution of type (b) (the ``interplanetary component''). The interplanetary
component usually peaks well ahead of the sunspot minimum and the amplitude of
the peak seemed to be a good predictor of the next sunspot maximum. However, it
is to be noted that the assumption that the ``surplus'' contribution to $aa$
originates from the interplanetary component only is likely to be erroneous,
especially for stronger cycles. It is known that the number of large solar
eruptions shows no unique relation to $R$: in particular, for $R>100$ their
frequency may vary by a factor of 3 \citep[see Figure~15 in][]{Hathaway:LRSP}, so
in some years they may well yield a contribution to $aa$ that greatly exceeds
the minimum contribution. A case in point was the ``Halloween events'' of 2003,
that very likely resulted in a large false contribution to the derived
``interplanetary'' $aa$ index \citep{Hathaway:Halloween}. As a result, the
geomagnetic precursor method based on the separation of the interplanetary
component predicts an unusually strong cycle 24 ($R_m\sim 150$), in contrast to
most other methods, including Ohl's method and the polar field precursor, which
suggest a weaker than average cycle ($R_m\sim 80$--$90$). 

In addition to the problem of neatly separating the interplanetary contribution
to geomagnetic disturbances, it is also wrong to assume that this interplanetary
contribution is dominated by the effect of polar magnetic fields at all times
during the cycle.  Indeed, \cite{Wang+Sheeley:geomg.precursor} point out that
the interplanetary magnetic field amplitude at the Ecliptic is related to the
equatorial dipole moment of the Sun that does not survive into the next cycle,
so despite its more limited practical use, Ohl's original method, based on the
minima of the $aa$ index is physically better founded, as the polar dipole
dominates around the minimum.  The total amount of open interplanetary flux,
more closely linked to polar fields, could still be determined from geomagnetic
activity if the interplanetary contribution to it is further split
into
\begin{descr} 
\item{(b1)} A contribution due to the varying solar wind speed (or to the
interplanetary magnetic field strength anticorrelated with it), which in turn
reflects the strength of the equatorial dipole. 
\item{(b2)} Another contribution due to the overall interplanetary field
strength or open magnetic flux, which ultimately reflects the axial dipole.
\end{descr} 
Clearly, if the solar wind speed contribution (b1) could also subtracted, a
physically better founded prediction method should result.  While in situ
spacecraft measurements for the solar wind speed and the interplanetary magnetic
field strength do not have the necessary time coverage, 
\cite{Svalgaard+:LDV1,Svalgaard+:LDV2}  and \cite{Rouillard+:openflux} devised a
method to reconstruct the variations of both variables from geomagnetic
measurements alone. Building on their results,
\cite{Wang+Sheeley:geomg.precursor} arrive at a prediction of $R_m=97\pm 25$ for
the maximum amplitude of solar cycle~24. To what extent the effect of the
Halloween 2003 events has been removed from this analysis is unclear. In any
case, the prediction agrees fairly well with that of \cite{Bhatt+} who, assuming
a preliminary minimum time of August 2008 and applying a modified form of Ohl's
method, predict a cycle maximum in late 2012, with an amplitude of $93\pm 20$.

The actual run of cycle~24 will be certainly most revealing from the point of
view of these complex interrelationships.

The open magnetic flux can also be derived from the extrapolation of solar
magnetograms using a potential field source surface model. The magnetograms
applied for this purpose may be actual observations or the output from surface
flux transport models, using the sunspot distribution (butterfly diagram) and
the meridional flow as input. Such models indicate that the observed latitude
independence of the interplanetary field strength (``split monopole'' structure)
is only reproduced if the source surface is far enough ($>10\,R_\odot$) and the
potential field model is modified to take into account the heliospheric current
sheet (current sheet source surface model,
\citealp{Schussler+:openflux, Jiang+:openflux}). The extrapolations
are generally found to agree well with in situ measurements where
these are available.

\subsection{Flows in the photosphere}
\label{sect:flows}

In the currently widely popular flux transport dynamo models the strong polar
fields prevalent around sunspot minimum are formed by the advection of following
polarity flux from active regions by the poleward meridional flow. Changes in
this flow may thus influence the would-be polar fields and thereby may serve as
precursors of the upcoming cycle.

Such changes, on the other hand, are also associated with the normal course of
the solar activity cycle, the overall flow at mid-latitudes being slower before
and during maxima and faster during the decay phase. Therefore, it is just the
cycle-to-cycle variation in this normal pattern that may be associated with the
activity variations between cycles. In this respect it is of interest to note
that the poleward flow in the late phases of cycle~23 seems to have had an
excess speed relative to the previous cycle \citep{Hathaway+Rightmire}. If this
were a latitude-independent amplitude modulation of the flow, then most flux
transport dynamo models would predict a stronger than average polar field at the
minimum, contrary to observations. On the other hand, in the surface flux
transport model of \cite{Wang+:varying.circ} an increased poleward flow actually
results in weaker polar fields, as it lets less leading polarity flux to diffuse
across the equator and cancel there. As the recent analysis by \cite{Munozjara}
has shown, the discrepancy resulted from the neglect of leading polarity flux in
the Babcock--Leighton source term in flux transport dynamo models, and it can be
remedied by substituting a pair of opposite polarity flux rings as source term
instead of the $\alpha$-term. With this correction, 2D flux transport and
surface flux transport models agree in predicting a weaker polar field for
faster meridional flow.

It is known from helioseismology that meridional flow speed fluctuations follow
a characteristic latitudinal pattern associated with torsional oscillations and
the butterfly diagram, consisting of a pair of axisymmetric bands of latitudinal
flows converging towards the activity belts, migrating towards the equator, and
accompanied by similar high-latitude poleward branches. This suggests 
interpreting the unusual meridional flow speeds observed during cycle~23 as an
increased amplitude of this migrating modulation, rather than a change in the
large-scale flow speed \citep{Cameron+:circ.belts}. In this case, the flows
converging on the activity belts tend to inhibit the transport of following
polarities to the poles, again resulting in a lower than usual polar field, as
observed (\citealp{Jiang+:merid.flow}; note, however, that \citealp{Svanda} find
no change in the flux transport in areas with increased flows). It is
interesting to note that the torsional oscillation pattern, and thus presumably
the associated meridional flow modulation pattern, was shown to be fairly well
reproduced by a microquenching mechanism due to magnetic flux emerging in the
active belts \citep{Petrovay+Dajka:torso}. Observational support for this notion
has been provided by the seismic detection of locally increased flow modulation
near active regions \citep{Svanda}. This suggests that stronger cycles may be
associated with a stronger modulation pattern, introducing a nonlinearity into
the flux transport dynamo model, as suggested by \cite{Jiang+:merid.flow}.

In addition to a variation in the amplitude of migrating flow modulations, their
migration speed may also influence the cycle. \cite{Howe+:torso.precursor} point
out that in the current minimum the equatorward drift of the torsional
oscillation shear belt corresponding to the active latitude of the cycle has
been slower than in the previous minimum. They suggest that this slowing may
explain the belated start of cycle~24.


\newpage

\section{Extrapolation Methods}
\label{sec:Extrapolation-Methods}

In contrast to precursor methods, extrapolation methods only use the time series
of sunspot numbers (or whichever solar activity indicator is considered) but
they generally rely on more than one previous point to identify trends that can
be used to extrapolate the data into the future. They are therefore also known
as \textit{time series analysis}, or, for historic reasons, \textit{regression
methods}.

A cornerstone of time series analysis is the assumption that the time series is
\textit{homogeneous}, i.e., the mathematical regularities underlying its
variations are the same at any point of time. This implies that a forecast for,
say, three years ahead has equal chance of success in the rising or decaying
phase of the sunspot cycle, across the maximum, or, in particular, across the
minimum. In this case, distinguishing intracycle and intercycle memory effects,
as we did in Sections \ref{sect:memory} and \ref{sect:precursor}, would be
meaningless. This concept of solar activity variations as a continuous process
stands in contrast to that underlying precursor methods, where solar cycles are
thought of as individual units lasting essentially from minimum to minimum,
correlations within a cycle being considerably stronger than from one cycle to
the next. While, as we have seen, there is significant empirical evidence for
the latter view, the possibility of time homogeneity cannot be discarded out of
hand. Firstly, if we consider the time series of global parameters (e.g.,
amplitudes) of cycles, homogeneity may indeed be assumed fairly safely. This
approach has rarely been used for the directly observed solar cycles as their
number is probably too low for meaningful inferences -- but the long data sets
from cosmogenic radionuclides are excellent candidates for time series analysis.

In addition, there may be good reasons to consider the option of homogeneity of
solar activity data even on the scale of the solar cycle.  Indeed, in dynamo
models the solar magnetic field simply oscillates between (weak) poloidal and
(strong) toroidal configuration: there is nothing inherently special about
either of the two, i.e., there is no \textit{a priori} reason to attribute a
special significance to solar minimum. While at first glance the butterfly
diagram suggests that starting a new cycle at the minimum is the only meaningful
way to do it, there may be equally good arguments for starting a new cycle at
the time of polar reversal. There is, therefore, plenty of motivation to try and
apply standard methods of time series analysis to sunspot data. 

Indeed, as the sunspot number series is a uniquely homogeneous and long data
set, collected over centuries and generated in a fairly carefully controlled
manner, it has become a favorite testbed of time series analysis methods and is
routinely used in textbooks and monographs for illustration purposes.
\citep{Box+Jenkins, Wei, Tong}. This section will summarize
the various approaches, proceeding, by and large, from the simplest towards the
most complex.

\subsection{Linear regression}
\label{sect:regression}

Linear (auto)regression means representing the value of a time series
at time $t$ by a linear combination of values at times $t-\Delta t$, $t-2\Delta
t$, $\dots$, $t-p\Delta t$.  Admitting some random error $\epsilon_n$, the value
of $R$ in point $n$ is
\[
R_n={R_0}+\sum_{i=1}^p c_{n-i} R_{n-i} +\epsilon_n
\]
where $p$ is the order of the autoregression and the $c_i$'s are weight
parameters. A further twist on the model admits a propagation of errors from the
previous $q$ points:
\[
R_n={R_0}+\sum_{i=1}^p c_{n-i} R_{n-i} +\epsilon_n +\sum_{i=1}^q
d_{n-i}\epsilon_{n-i}
\]
This known as the ARMA (AutoRegressive Moving Average) model.

Linear regression techniques have been widely used for solar activity prediction
during the course of an ongoing cycle. Their application for cycle-to-cycle
prediction has been less common and successful \citep{Lomb+Andersen,
  Box+Jenkins, Wei}.

\cite{Brajsa+:gleissbg} applied an ARMA model to the series of annual values
of $R$. A successful fit was found for $p=6$, $q=6$. Using this fit, the next
solar maximum was predicted to take place around 2012.0 with an amplitude 
$90\pm 27$, and the following minimum occurring in 2017.

Instead of applying an autoregression model directly to SSN data,
\cite{Hiremath} applied it to a forced and damped harmonic oscillator model
claimed to well represent the SSN series. This resulted in a predicted amplitude
of $110\pm 10$ for solar cycle 24, with the cycle starting in mid-2008 and
lasting 9.34 years.


\subsection{Spectral methods}
\label{sect:spectral}

\begin{nquote}
{\sl ...the use of any mathematical algorithm to derive hidden periodicities
from the data always entails the question as to whether the resulting cycles are
not introduced either by the particular numerical method used or by the time
interval analyzed.}\\
\strut\hfill\citep{deMeyer:impulse}
\end{nquote}

Spectral analysis of the sunspot number record is used for prediction under the
assumption that the main reason of variability in the solar cycle is a long-term
modulation due to one or more periods.  

The usual approach to the problem is the purely formal one of representing the
sunspot record with the superposition of eigenfunctions forming an orthogonal
basis. From a technical point of view, spectral methods are a complicated form
of linear regression. The analysis can be performed by any of the widely used
means of harmonic analysis:

(1) {Least squares (LS) frequency analysis} (sometimes called ``Lomb--Scargle
periodogram'') consists in finding by trial and error the best fitting sine
curve to the data using the least squares method, subtracting it
(``prewhitening''), then repeating the procedure until the residuals become
indistinguishable from white noise. The first serious attempt at sunspot cycle
prediction, due to \cite{Kimura}, belonged to this group. The analysis resulted
in a large number of peaks with dubious physical significance. The prediction
given for the upcoming cycle~15 failed, the forecasted amplitude being $\sim 60$
while the cycle actually peaked at 105. However, it is interesting to note that
Kimura correctly predicted the long term strengthening of solar activity during
the first half of the 20th century! LS frequency analysis on sunspot data was
also performed by \cite{Lomb+Andersen}, with similar results for the spectrum.

(2) \textit{Fourier analysis} is probably the most commonly used method of spectral
decomposition in science. It has been applied to sunspot data from the beginning
of the 20th century \citep{Turner1, Turner2, Michelson}.
\cite{Vitinsky:book2} judges Fourier-based forecasts even less reliable than LS
periodogram methods. Indeed, for instance \cite{Cole} predicted cycle~21 to
have a peak amplitude of 60, while the real value proved to be nearly twice
that. 

(3) The \textit{maximum entropy method (MEM)} relies on the Wiener--Khinchin
theorem that the power spectrum is the Fourier transform of the autocorrelation
function. Calculating the autocorrelation of a time series for $M\ll N$ points
and extrapolating it further in time in a particular way to ensure maximal
entropy can yield a spectrum that extends to arbitrarily low frequencies despite
the shortness of the data segment considered, and also has the property of being
able to reproduce sharp spectral features (if such are present in the data in
the first place). A good description of the method is given by \cite{Ables},
accompanied with some propaganda for it -- see \cite{numrec} for a more balanced
account of its pros and cons. The use of MEM for sunspot number prediction was
pioneered by \cite{Currie:max.entropy}.  Using maximum entropy method combined
with multiple regression analysis (MRA) to estimate the amplitudes and phases,
\cite{Kane:MEM24} arrived at a prediction of 80 to 101 for the maximum amplitude
of cycle~24. It should be noted that the same method yielded a prediction
\citep{Kane:MEM23} for cycle~23 that was far off the mark.

(4) \textit{Singular spectrum analysis (SSA)} is a relatively novel method for the
orthogonal decomposition of a time series. While in the methods discussed above
the base was fixed (the trigonometric functions), SSA allows for the
identification of a set of othogonal eigenfunctions that are most suitable for
the problem. This is done by a principal component analysis of the covariance
matrix $r_{ik}=\langle R_i R_{i+k}\rangle$. SSA was first applied to the sunspot
record by \cite{Rangarajan:SSA} who only used this method for pre-filtering
before the application of MEM. \cite{Loskutov:SSA} who also give a good
description of the method, already made a prediction for cycle~24: a peak
amplitude of 117. More recently, the forecast has been corrected slightly
downwards to 106 \citep{Kuzanyan:predict.poster}.

The dismal performance of spectral predictions with the methods (1)--(3)
indicates that the sunpot number series cannot be  well represented by the
superposition of a limited number of fixed periodic components. Instead,

\begin{itemize}
\item the periods may be time dependent
\item the system may be quasiperiodic, with a significant finite width of the
periodic peaks (esp.\ the 11-year peak)
\item there may be non-periodic (i.e., chaotic or stochastic) components in the
behaviour of the system, manifested as a continuous background in the spectrum.
\end{itemize}

In practice, all three effects suggested above may play some part. The first
mentioned effect, time dependence, may in fact be studied within the framework
of spectral analysis. MEM and SSA are intrinsically capable of detecting or
representing time dependence in the spectrum, while LS and Fourier analysis can
study time dependence by sliding an appropriate data window across the period
covered by observations. If the window is Gaussian with a width proportional to
the frequency we arrive at the popular \textit{wavelet analysis}.  This method
was applied to the sunspot number series by \cite{Ochadlick+},
\cite{Vigouroux+}, \cite{Frick+:wavelet}, \cite{Fligge+} and \cite{Li+:wavelet}
who could confirm the existence and slight variation of the 11-year cycle
and the Gleissberg-cycle. Recently, \cite{Kollath+Olah:1} called attention to a
variety of other generalized time dependent spectral analysis methods, of which
the pseudo-Wigner transform yields especially clear details
(cf.~Figure~\ref{fig:Kollath}). The time varying character of the basic periods
makes it difficult to use these results for prediction purposes but they are
able to shed some light on the variation as well as the presistent or
intermittent nature of the periods determining solar activity. 

In summary, it is fair to say that forecasts based on harmonic analysis are
notoriously unreliable.  The secular variation of the basic periods, obeying as
yet unknown rules, would render harmonic analysis practically useless for the
prediction of solar cycles even if solar activity could indeed be described by a
superposition of periodic functions. Although they may be potentially useful for
very long term prediction (on centennial scales), when it comes to
cycle-to-cycle forecasts the best we can hope from spectral studies is
apparently an indirect contribution, by constraining dynamo models with the
inambiguously detected periodicities. 

In what remains from this subsection, we briefly review what these apparently
physically real periods are and what impact they may have on solar cycle
prediction.

\epubtkImage{}{%
  \begin{figure}[htbp]
\centerline{\includegraphics[width=0.9\textwidth]{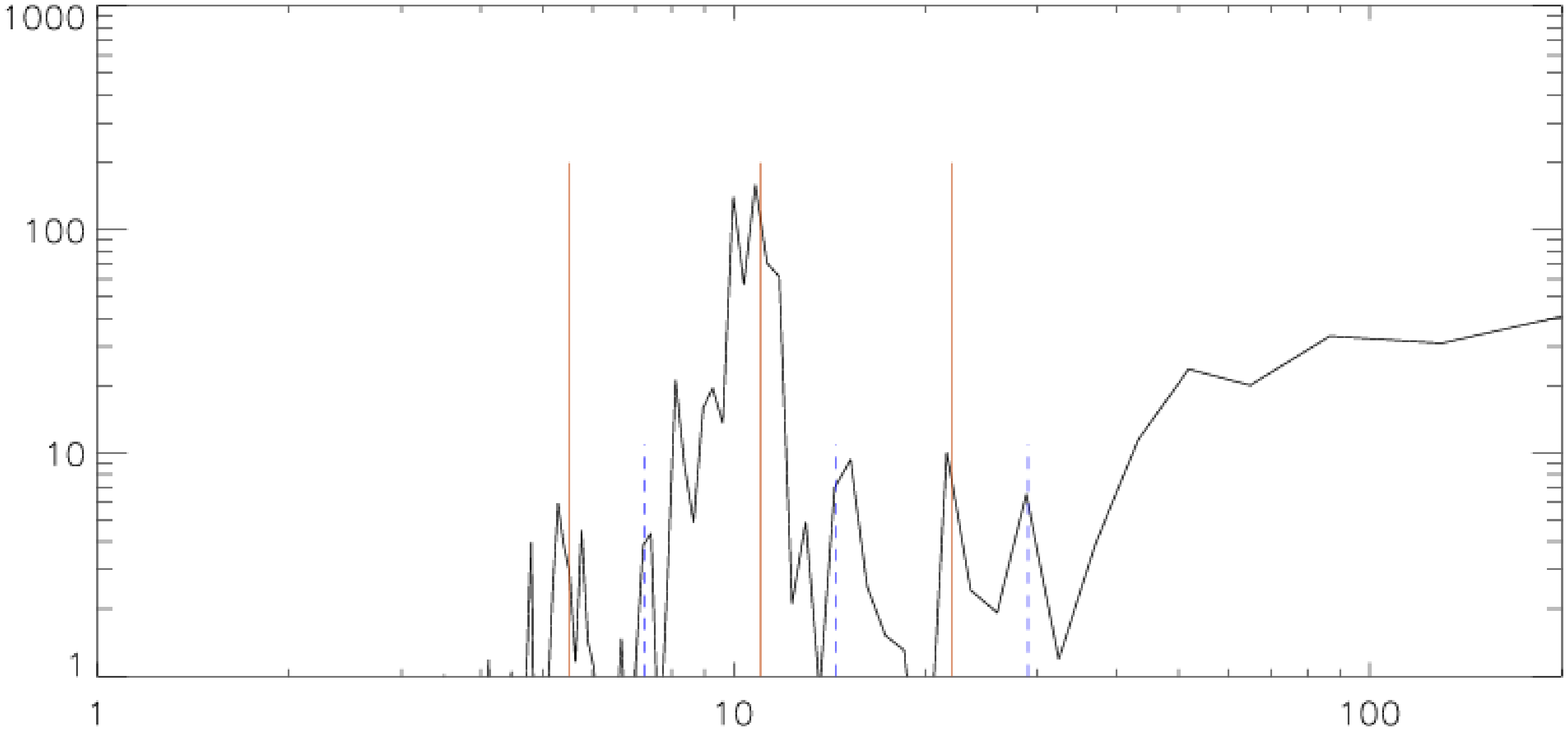}}
\caption{Power spectrum of the smoothed monthly sunspot number series for the
period 1749--2008. Solid vertical bars mark the 11-year period, its first
harmonic and subharmonic; dashed vertical bars are drawn at a fiducial period of
14.5 years, its harmonic and subharmonic.}
\label{fig:ssn_powsp0}
\end{figure}}

\subsubsection{The 11-year cycle and its harmonics}

As an example of the period spectrum obtained by these methods, in
Figure~\ref{fig:ssn_powsp0} we present the FFT based power spectrum
estimate of the smoothed sunspot number record. Three main features
are immediately noticed:

\begin{itemize}
\item The dominant 11-year peak, with its sidelobes and its
5.5-year harmonic 
\item The 22-year subharmonic, representing the even--odd rule 
\item The significant power present at periods longer than 50~years,
associated with the Gleissberg cycle
\end{itemize}

The dominant peak in the power spectrum is at $\sim 11$ years. Significant power
is also present at the first harmonic of this period, at 5.5~years. This is
hardly surprising as the sunspot number cycles, as presented in
Figure~\ref{fig:SSNrecord}, have a markedly asymmetrical profile. It is a
characteristic of Fourier decomposition that in any periodic series of cycles
where the profiles of individual cycles are non-sinusoidal, all harmonics of the
base period will appear in the spectrum. 

Indeed, were it not for the 13-month smoothing, higher harmonics could also be
expected to appear in the power spectrum. It has been proposed
\citep{Krivova+Solanki:1.3yr} that these harmonics are detected in the sunspot
record and that they may be related to the periodicities of $\sim 1.3$ years
intermittently observed in solar wind speed \citep{Richardson+:1.3yr,
  Paularena+:1.3yr, Szabo+:1.3yr, Mursula+Zieger:1.3yr,
  Lockwood:1.3yr} and in the internal rotation velocity of the Sun
\citep[][Sect.~10.1]{Howe:LRSP}. An analoguous intermittent 2.5 year
variation in the solar neutrino flux \citep{Shirai} may also belong to
this group of phenomena. It may be worth noting that, from the other
end of the period spectrum, the 154-day Rieger period in solar flare
occurrence \citep{Rieger+, Bai+Cliver} has also been tentatively
linked to the 1.3-year periodicity. Unusually strong excitation of
such high harmonics of the Schwabe cycle may possibly be explained by
excitation due to unstable Rossby waves in the tachocline
\citep{Zaqarashvili+}.

The 11-year peak in the power spectrum has substantial width, related to the
rather wide variation in cycle lengths in the range 9--13 years. Yet
Figure~\ref{fig:ssn_powsp0} seems to suggest the presence of a well detached
second peak in the spectrum at a period of $\sim 14$ years. The presence of a
distinct peak at the first harmonic and even at the subharmonic of this period
seems to support its reality. Indeed, peaks at around 14 and 7 years were
already found by other researchers \citep[e.g.,][]{Kimura,
  Currie:max.entropy} who suggested that these may be real secondary
periods of sunspot activity. 

The situation is, however, more prosaic. Constraining the time interval
considered to data more recent than 1850, from which time the sunspot number
series is considered to be more reliable, the 14.5-year secondary peak and its
harmonics completely disappear. On the other hand, the power spectrum for the
years 1783--1835 indicates that the appearance of the 14.5-year secondary peak
in the complete series is almost entirely due to the strong predominance of this
period (and its harmonic) in that interval. This interval covers the unusually
long cycle~4 and the Dalton minimum, consisting of three consecutive unusually
weak cycles, when the ``normal'' 11-year mode of operation was completely
suppressed.

As pointed out by \cite{Py:Rio}, this probably does not imply that the Sun was
operating in a different mode during the Dalton minimum, the cycle length being
14.5 years instead of the usual 11 years. Instead, the effect may be explained
by the well known inverse correlation between cycle length and amplitude, which
in turn is the consequence of the strong inverse correlation between rise rate
and cycle amplitude (Waldmeier effect), combined with a much weaker or
nonexistent correlation between decay rate and amplitude (see
Section~\ref{sect:Waldmeier}). The cycles around the
Dalton minimum, then, seem to lie at the low amplitude (or long period) end of a
continuum representing the well known cycle length--amplitude relation,
ultimately explained by the Waldmeier effect. 

A major consequence of this is that the detailed distribution of peaks
varies significantly depending on the interval of time considered. Indeed,
\cite{Kollath+Olah:1} recently applied time dependent harmonic analysis to the
sunspot number series and found that the dominant periods have shown 
systematic secular changes during the past 300~years
(Figure~\ref{fig:Kollath}). For instance, the basic period seems to
have shortened from 11~years to 10~years between 1850 and 1950, with
some moderate increase in the last 50~years. (This is consistent with
the known anticorrelation between cycle length and amplitude,
cf.\ Section~\ref{sect:Waldmeier}.) 

\epubtkImage{}{%
  \begin{figure}[htbp]
\centerline{\includegraphics[width=0.6\textwidth]{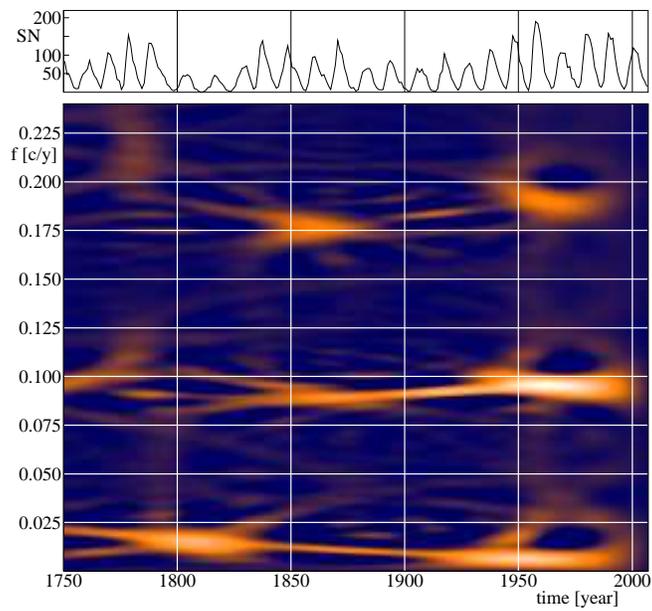}}
\caption{Pseudo-Wigner power distribution in the sunspot number record, with
time on the abscissa and frequency on the ordinate. The three horizontal bands
of high power correspond, from bottom to top, to the Gleissberg cycle, the
11-year cycle and its first harmonic. The sunspot number curve is shown on top
for guidance. (Figure courtesy Z.~Koll\'ath)}
\label{fig:Kollath}
\end{figure}}

\subsubsection{The even--odd (a.k.a.\ Gnevyshev--Ohl) rule}
\label{sect:evenodd}

A cursory look at Figure~\ref{fig:SSNrecord} shows that solar cycles often
follow an alternating pattern of higher and lower maxima. In this apparent
pattern, already noticed by the early observers \citep[e.g,][]{Turner:evenodd},
odd cycles have been typically stronger than even cycles in the last two
centuries.

This even--odd rule can be given two interpretations: a ``weak'' one of a general
tendency of alternation between even and odd cycles in amplitude, or a
``strong'' one of a specific numerical relation between the amplitudes of
consecutive cycles. 

Let us first consider the rule in its weak interpretation. At first sight the
rule admits many exceptions, but the amplitude of solar cycles depends on the
particular measuring method used. Exceptions from the even--odd alternation rule
become less common if a long term trend (calculated by applying a 12221 or 121
filter, see Section~\ref{sect:secular}) is subtracted from the data
\citep{Charbonneau:evenodd}, or if integrated cycle amplitudes (sums of annual
mean sunspot numbers during the cycle) are used \citep{Gnevyshev+Ohl}. 

In fact, as evident from, e.g., the work of \cite{Mursula+:evenodd} where cycle
amplitudes are based on group sunspot numbers and the amplitude of a cycle is
defined as the sum of the annual GSN value over the course of the cycle, the
odd--even alternation may be considered as \textit{strictly} valid
with only four exceptions:

\begin{itemize}
\item In the pairs 7--8 and 17--18, odd cycles are followed by stronger even
cycles at the end of Dalton minimum and at the beginning of the Modern Maximum.
These exceptions could be made to disappear by the subtraction of the long term
trend as suggested by \cite{Charbonneau:evenodd}. 
\item The pair 22--23 represents another apparent break of the weak even--odd
rule which is not easily explained away, even though the relative difference is
smaller if the Kislovodsk sunspot number series is used
\citep{Nagovitsyn+:evenodd}. The possibility is obviously there that the
subtraction of the long term trend may resolve the problem but we have no way
to tell in the near future.
\item Prior to cycle~5, the phase of the alternation was opposite, even cycles
being stronger than odd cycles.  As cycle~4 is known to have been anomalously
long anyway (the so-called ``phase catastrophe'' in the solar cycle,
\citealp{Vitinsky+:book}) and its decaying phase is not well covered by
observations \citep{Vaquero}, this gave rise to the suggestion of a ``lost solar
cycle'' between cycles~4 and 5 \citep{Usoskin+:lostcycle, Usoskin+Mursula}. This
cycle, however, would have been even more anomalous than cycle 4 and despite
intensive searches in historic data the evidence is still not quite conclusive
(\citealp{Krivova+:lostcycle}; see, however,
\citealp{Usoskin+:lostcycle.butterfly}).
\end{itemize}

The issue whether the even--odd rule can go through phase jumps or not is
important with respect to its possible origin. One plausible possibility is that
the alternation is due to the superposition of a steady primordial  magnetic
field component on the oscillatory magnetic field generated by the dynamo
\citep{Levy+Boyer}. In this case, any phase jump in the Gnevyshev--Ohl rule
should imply a phase jump in Hale's polarity rules, too.  Alternatively,
persistent even--odd alternation may also arise in nonlinear dynamos as a
period--2 limit cycle \citep{Durney:evenodd}; with a stochastic forcing
occasional phase jumps are also possible \citep{Charbonneau:evenodd,
  Charbonneau+:evenodd}.

While we have no information on this from the 18th century phase jump, we can be
certain that there was no such phase jump in polarities in the last two decades,
even though the even--odd rule seems to have been broken again. It will be
interesting to see when (and if) the even--odd rule settles in again, whether it
will have done so with a phase jump or not. For instance, if cycle~25 will again exceed
cycle~24 it would seem that no phase jump occurred and both theoretical  options
are still open. But if cycle~25 will represent a further weakening from
cycle~24, followed by a stronger cycle~26, a phase jump will have occurred,
which may exclude the primordial field origin of the rule if Hale's polarity
rules remain unchanged.

Let us now discuss the stronger interpretation of the even--odd rule. In the
first quantitative study of the relative amplitudes of consecutive cycles,
\cite{Gnevyshev+Ohl} found a rather tight correlation between the time
integrated amplitudes of even and subsequent odd cycles, while the correlation
between odd cycles and subsequent even cycles was found to be much less strong.
This gave rise to the notion that solar cycles come in ``two-packs'' as
even--odd pairs. \cite{Nagovitsyn+:evenodd} confirmed this puzzling finding on
the basis of data covering the whole period of telescopic observations (and
renumbering cycles before 1790 in accordance with the lost cycle hypothesis);
they also argue that cycle pair 22--23 does not deviate strongly from the
even--odd correlation curve so it should not be considered a ``real'' exception
to the even--odd rule.

The fact that shortly after its formulation by \cite{Gnevyshev+Ohl}, the
(strong) even--odd rule was used by \cite{Kopecky:pred19} to successfully
predict the unusually strong cycle~19 made this method particularly popular for
forecast purposes. However, forecasts based on the even--odd rule completely 
failed for cycle~23, overpredicting the amplitude by $>50\%$
\citep[see review by][]{Li+:pred.rev}. Taken together with the
implausibility of the suggested two-pack system, this shows that it is
probably wiser to take the position that ``extraordinary claims need
extraordinary evidence'' -- which is yet to be provided in the case of
the ``strong'' even--odd rule.

Finally, in the context of the even--odd rule, it is also worth mentioning the
three-cycle regularity proposed by \cite{Ahluwalia:pred23}. Even though the
evidence presented for the alleged triadic pattern is not overwhelming, this
method resulted in one of the few successful predictions for the amplitude of
cycle 23. 

\subsubsection{The Gleissberg cycle}

Besides the changes in the length of the 11-year cycle related to the
amplitude--cycle length correlation, even more significant are the variations in
the period of the so-called Gleissberg cycle \citep{Gleissberg:cycle}. This
``cycle'', corresponding to the 60--120 year ``plateau'' in
Figure~\ref{fig:ssn_powsp0} was actually first noticed by Wolf, who placed it in
the range 55--80 years \citep[see][for a discussion of the history of the
studies of the Gleissberg cycle]{Richard:Gleissberg}. Researchers in the middle
of the 20th century characterized it as an 80--100 year variation.
Figure~\ref{fig:Kollath} explains why so widely differing periods were found in
different studies: the period has in fact shown a secular increase in the past
300 years, from about 50 years in the early 18th century, to a current value
exceeding 140 years. This increased length of the Gleissberg cycle also agrees
with the results of \cite{FDE+:midtermvar}.

The detection of $\sim 100$ year periods in a data set of 300 years is of course
always questionable, especially if the period is even claimed to be varying. 
However, the very clear, and, most importantly, nearly linear secular trend seen
in Figure~\ref{fig:Kollath} argues convincingly for the reality of the period in
question. This clear appearance of the period is due to the carefully optimized
choice of the kernel function in the time--frequency analysis, a method
resulting in a so-called pseudo-Wigner distribution (PWD). In addition, in their
study \cite{Kollath+Olah:1} present an extremely conscientious test of the
reliability of their methods, effectively proving that the most salient features
in their PWD are not artefacts. (The method was subsequently also applied to
stellar activity, \citealp{Kollath+Olah:2}.)  This is the most compelling evidence
for the reality of the Gleissberg cycle yet presented.

\subsubsection{Supersecular cycles}

For the 210-year Suess cycle, \cite{McCracken+Beer:ICRC} present further
evidence for the temporally intermittent nature of this marked peak in the
spectrum of solar proxies. The Suess cycle seems to have a role in regulating
the recurrence rate of grand minima. Grand minima, in turn, only seem to occur
during $<1$ kiloyear intervals (``Sp\"orer events'') around the minimum of the
$\sim 2400$-year Hallstatt cycle.

For further discussion of long term variations in solar activity we refer the
reader to the reviews by \cite{Beer+:SSR} and \cite{Usoskin:LRSP}.

\subsection{Nonlinear methods}
\label{sect:nonlin}

\begin{nquote}
{\sl ``...every complicated question has a simple answer which is wrong. Analysing a
time series with a nonlinear approach is definitely a complicated problem.
Simple answers have been repeatedly offered in the literature, quoting numerical
values for attractor dimensions for any conceivable system.''}\\
\strut\hfill\citep{Hegger+}
\end{nquote}

The nonlinearities in the dynamo equations readily give rise to chaotic
behaviour of the solutions. The long term behaviour of solar activity, with
phenomena like grand minima and grand maxima, is also suggestive of a chaotic
system. While chaotic systems are inherently unpredictable on long enough time
scales, their deterministic nature does admit forecast within a limited range.
It is therefore natural to explore this possibility from the point of view of
solar cycle prediction.

\subsubsection{Attractor analysis and phase space reconstruction: the
  pros ...}
\label{sect:nldpro}

Assuming that the previous $(M-1)$ values of the sunspot number do in some way
determine the current expected value, our problem becomes restricted to an
$M$-dimensional \textit{phase space}, the dimensions being the current value and
the $(M-1)$ previous values. With a time series of length $N$, we have $N-M+1$
points fixed in the phase space, consecutive points being connected by a line.
This phase space trajectory is a sampling of the \textit{attractor} of the physical
system underlying the solar cycle (with some random noise added to it). The
attractor represents a \textit{mapping} in phase space which maps each point into
the one the system occupies in the next time step: if this mapping is known to
a good approximation, it can be used to extend the trajectory towards the
future.

For the mapping to be known, $M$ needs to be high enough to avoid self-crossings
in the phase space trajectory (otherwise the mapping is not unique) but low
enough so that the trajectory still yields a good sampling of the attractor. The
lowest integer dimension satisfying these conditions is the \textit{embedding
dimension} $D$ of the attractor (which may have a fractal dimension itself).

Once the attractor has been identified, its mathematical description may be done
in two ways.

(1) \textit{Parametric fitting} of the attractor mapping in phase space. The simplest
method is the  piecewise linear fit suggested by \cite{Farmer+Sidorowich} and
applied in several solar prediction attempts, e.g., \cite{Kurths+Ruzmaikin}.
Using a method belonging to this group, for cycle~24 \cite{Kilcik+} predict a
peak amplitude of 87 to be reached in late 2012. Alternatively, a global
nonlinear fit can also be used: this is the method applied by \cite{Serre+Nesme}
as the first step in their global flow reconstruction (GFR) approach. 

(2) \textit{Nonparametric fitting.} The simplest nonparametric fit is to find the
closest known attractor point to ours (in the $(M-1)$-dimensional subspace
excluding the last value) and then using this for a prediction, as done  by
\cite{Jensen}. (This resulted in so large random forecast errors that it is
practically unsuitable for prediction.) \textit{Neural networks,} discussed in more
detail in Section~\ref{sect:neural} below, are a much more sophisticated
nonparametric fitting device.

(3) Indirectly, one may try to find a set of differential equations describing a
system that gives rise to an attractor with properties similar to the observed.
In this case there is no guarantee that the derived equations will be unique, as
an alternative, completely different set may also give rise to a very similar 
attractor. This arbitrariness of the choice is not necessarily a problem from
the point of view of prediction as it is only the mapping (the attractor
structure) that matters. Such phase space reconstruction by a set of governing
equations was performed, e.g., by \cite{Serre+Nesme} or \cite{Aguirre+}; for cycle
24 the latter authors predict a peak amplitude of $65\pm 16$. On the other hand,
instead of putting up with any arbitrary set of equations correctly reproducing
the phase space, one might make an effort to find a set with a structure
reasonably similar to the dynamo equations so they can be given a
meaningful physical interpretation. Methods following this latter approach will
be discussed in Sections~\ref{sect:truncated} and \ref{sect:oscillator}.

\subsubsection{... the cons ...}

Finding the embedding dimension and the attractor structure is not a trivial
task, as shown by the widely diverging results different researchers arrived at.
One way to find the correct embedding dimension is the false nearest neighbours
method \citep{Kennel+}, essentially designed to identify self-crossing in the
phase space trajectory, in which case the dimension $M$ needs to be increased.
But self-crossings are to some extent inevitable, due to the stochastic
component superimposed on the deterministic skeleton of the system. 

As a result, the determination of the minimal necessary embedding dimension is
usually done indirectly. One indirect method fairly popular in the solar
community is the approach proposed by \cite{Sugihara+May} where the correct
dimension is basically figured out on the basis of how successfully the model,
fit to the first part of the data set, can ``predict'' the second part (using a
piecewise linear mapping).

Another widely used approach, due to \cite{Grassberger+Procaccia}, starts by
determining the correlation dimension of the attractor, by simply counting how
the number of neighbours in an embedding space of dimension $M\gg 1$ increases
with the distance from a point. If the attractor is a lower dimensional manifold
in the embedding space and it is sufficiently densely sampled by our data 
then the logarithmic steepness $d$ of this function should be constant over a
considerable stretch of the curve: this is the correlation dimension $d$. Now,
we can increase $M$ gradually and see at what value $d$ saturates: that value
determines the attractor dimension, while the value of $M$ where saturation is
reached yields the embedding dimension. 

The first nonlinear time series studies of solar activity indicators suggested a
time series spacing of 2--5 years, an attractor dimension $\sim 2$--$3$ and an
embedding dimension of 3--4 \citep{Kurths+Ruzmaikin, Gizzatullina+}. Other
researchers, however, were unable to confirm these results, either reporting
very different values or not finding any evidence for a low dimensional
attractor at all \citep{Calvo+, Price+:nochaos, Carbonell+:nochaos, Kilcik+,
Hanslmeier+Brajsa}. In particular, I would like to call attention to the paper
by \cite{Jensen}, which, according to ADS and WoS, has received a grand total of
zero citations (!) up to 2010, yet it displays an exemplary no-nonsense approach
to the problem of sunspot number prediction by nonlinear time series methods.
Unlike so many other researchers, the author of that paper does not fool himself
into believing to see a linear segment on the logarithmic correlation integral
curve (his Figure~4); instead, he demonstrates on a simple example that the
actual curve can be perfectly well reproduced by a simple stochastic process.

These contradictory results obviously do not imply that the mechansim generating
solar activity is \textit{not} chaotic. For a reliable determination a long time
series is desirable to ensure a sufficiently large number of neighbours in a
phase space volume small enough compared to the global scale of the attractor.
Solar data sets (even the cosmogenic radionuclide proxies extending over
millennia but providing only a decadal sampling) are typically too short and
sparse for this. In addition, clearly distinguishing between the phase space
fingerprints of chaotic and stochastic processes is an unsolved problem of
nonlinear dynamics which is not unique to solar physics. A number of methods
have been suggested to identify chaos unambiguously in a time series but none of
them has been generally accepted and this topic is currently a subject of
ongoing research -- see, e.g., the work of \cite{Freitas+} which demonstrates that
the method of ``noise titration'', somewhat akin to the Sugihara--May algorithm,
is uncapable of distinguishing superimposed coloured noise from intrinsically
chaotic systems.

\subsubsection{... and the upshot}

Starting from the 1980s many researchers jumped on the chaos bandwagon, applying
nonlinear time series methods designed for the study of chaotic systems to a
wide variety of empirical data, including solar activity parameters. From the
1990s, however, especially after the publication of the influential book by
\cite{Kantz+Schreiber}, it was increasingly realized that the applicability of
these nonlinear algorithms does not in itself prove the predominantly chaotic
nature of the system considered. In particular, stochastic noise superposed on a
simple, regular, deterministic skeleton can also give rise to phase space
characteristics that are hard to tell from low dimensional chaos, especially if
strong smoothing is applied to the data. As a result, the pendulum has swung in
the opposite direction and currently the prevailing view is that there is no
clear cut evidence for chaos in solar activity data \citep{Panchev+Tsekov}.

One might take the position that any forecast based on attractor analysis is
only as good as the underlying assumption of a chaotic system is: if that
assumption is unverifiable from the data, prediction attempts are pointless.
This, however, is probably a too hasty judgment. As we will see, the potentially
most useful product of phase space reconstruction attempts is the inferences
they allow regarding the nature of the underlying physical system (chaotic or
not), even offering a chance to constrain the form of the dynamo equations
relevant for the Sun. As discussed in the next section, such truncated models
may be used for forecast directly, or alternatively, the insight they yield into
the mechanisms of the dynamo may be used to construct more sophisticated dynamo
models.

\subsubsection{Neural networks}
\label{sect:neural}

Neural networks are algorithms built up from a large number of small
interconnected units (``neurons'' or ``threshold logic units''), each of which
is only capable of performing a simple nonlinear operation on an input signal,
essentially described by a step function or its generalized (rounded) version, a
sigmoid function. To identify the optimal values of thresholds and weights
parameterizing the sigmoid functions of each neuron, an algorithm called ``back
propagation rule'' is employed which minimizes (with or without human guidance)
the error between the predicted and observed values in a process called
``training'' of the network. Once the network has been correctly trained, it is
capable of further predictions.

The point is that any arbitrary multidimensional nonlinear mapping may be
approximated by a combination of stepfunctions to a good degree -- so, as
mentioned in Section~\ref{sect:nldpro} above, the neural network can be used to
find the nonlinear mapping corresponding to the attractor of the given time
series. 

More detailed introductions to the method are given by \cite{Blais+Mertz} and by
\cite{Calvo+}; the latter authors were also the first to apply a neural network
for sunspot number prediction. Unfortunately, despite their claim of being able
to ``predict'' (i.e. postdict) some earlier cycles correctly, their prediction
for cycle~23 was off by a wide margin (predicted peak amplitude of 166 as
opposed to 121 observed). One of the neural network forecasts for cycle~24
\citep{Maris+Oncica} was equally far off, predicting a maximum of 145 as early
as December 2009, while another one \citep{Uwamahoro} yields a more
conservative value of $117.5\pm 8.5$ for 2012.


\newpage

\section{Model-Based Predictions}
\label{sec:Model-Based-Predictions}

\begin{nquote}
{\sl Progress in dynamo theory is extremely difficult, as it can be made only by
understanding the interaction of turbulent plasma motions with magnetic fields.
Indeed, the extreme conditions within the solar interior make this a formidable
task. ... Any predictions made with such models should be treated with extreme
caution (or perhaps disregarded), as they lack solid physical underpinnings.}\\
\strut\hfill{\citep{Tobias+:antiDikpati}}
\end{nquote}

While attempts to predict future solar cycles on the basis of the empirical
sunspot number record have a century-old history, predictions based on physical
models of solar activity only started a few years ago. The background of this
new trend is, however, not some significant improvement in our understanding of
the solar dynamo. Rather, it is the availability of increasingly fast new
computers that made it possible to fine-tune the parameters of certain dynamo
models to reproduce the available sunspot record to a good degree of accuracy
and to apply data assimilation methods (such as those used in terrestrial
weather prediction) to these models. This is not without perils. On the one
hand, the capability of multiparametric models to fit a multitude of
observational data does not prove the conceptual correctness of the underlying
model. On the other hand, in chaotic or stochastic systems such as the solar
dynamo, fitting a model to existing data will not lead to a good prediction
beyond a certain time span, the extent of which can only be objectively
assessed by ``postdiction'' tests, i.e. checking the models predictive skill by
trying to ``predict'' previous solar cycles and comparing those predictions to
available data. Apparently successful postdiction tests have led some groups to
claim a breakthrough in solar cycle prediction owing to the model-based
approach \citep{Dikpati+:prediction, Kitiashvili+:prediction}. Yet,
as we will see in the following discussion, a closer inspection of these claims
raises many questions regarding the role that the reliance on a particular
physical dynamo model plays in the success of their predictions.

\subsection{The solar dynamo: a brief summary of current models}

Extensive summaries of the current standing of solar dynamo theory are given in
the reviews by \cite{Petrovay:SOLSPA}, \cite{Ossendrijver:dynamo.review},
\cite{Charbonneau:livingreview}, and \cite{Solanki+:RPP}. As explained in detail
in those reviews, all current models are based on the mean-field theory approach
wherein a coupled system of partial differential equations governs the evolution
of the toroidal and poloidal components of the large-scale magnetic field. The
large-scale field is assumed to be axially symmetric in practically all current
models. In some nonlinear models the averaged equation of motion, governing
large-scale flows is also coupled into the system. 

In the simplest case of homogeneous and isotropic turbulence, where  the scale
$l$ of turbulence is small compared to the scale $L$ of the mean variables
(scale separation hypothesis), the dynamo equations have the form
\begin{equation}
  \pdv{\vc B}t=\nabla\times(\vc U\times\vc B+\alpha\vc B) 
  -\nabla\times(\eta_T\times\nabla\vc B)  \label{eq:dynamo}
\end{equation}
Here $\vc B$ and $\vc U$ are the large-scale mean magnetic field and flow
speed, respectively; $\eta_T$ is the magnetic diffusivity (dominated by the
turbulent contribution for the highly conductive solar plasma), while $\alpha$
is a parameter related to the non-mirror symmetric character of the magnetized
plasma flow. 

In the case of axial symmetry the mean flow $\vc U$ may be split into a
meridional circulation $\vc U_c$ and a differential rotation characterized by
the angular velocity profile $\Omega_0(r,\theta)$:
\[ \vc U=\vc U_c + r\sin\theta\,\Omega_0\,\vc e_\phi   \]
where $r$, $\theta$, $\phi$ are spherical coordinates and $\vc e_\phi$ is the
azimuthal unit vector. Now introducing the shear 
\[ {\vc\Omega}=r\sin\theta\,\nabla\Omega_0  , \qquad
\Omega=-\sgn{\dv{\Omega_0}r}\cdot|\vc\Omega| , \]
assuming $\alpha\ll\Omega L$ and ignoring spatial derivatives of $\alpha$ and
$\eta_T$, Equation~(\ref{eq:dynamo}) simplifies to the pair
\begin{equation}
  \pdv At=\alpha B-(\vc U_c\cdot\nabla)A-(\nabla\cdot\vc U_c)A+\eta_T\,\nabla^2 A 
  \label{eq:pol} 
\end{equation}
\begin{equation} \pdv Bt=\Omega\,\pdv Ax 
   -(\vc U_c\cdot\nabla)B -(\nabla\cdot\vc U_c)B
   +\eta_T\,\nabla^2 B     \label{eq:tor}  ,  
\end{equation}
where $B$ and $A$ are the toroidal (azimuthal) components of the magnetic field
and of the vector potential, respectively, and $\pdv Ax$ is to be evaluated in
the direction $90^\circ$ clockwards of $\vec\Omega$ (along the isorotation
surface) in the meridional plane. These are the classic $\alpha\Omega$ dynamo
equations, including a meridional flow.

In the more mainstream solar dynamo models the strong toroidal field is now
generally thought to reside near the bottom of the solar convective zone.
Indeed, it is known that a variety of flux transport mechanisms such as
pumping \citep{Petrovay:NATO} remove magnetic flux from the solar convective
zone on a timescale short compared to the solar cycle.  Following earlier
simpler numerical experiments, recent MHD numerical simulations have indeed
demonstrated this pumping of large scale magnetic flux from the convective zone
into the tachocline below, where it forms strong coherent toroidal fields
\citep{Browning+:dynsimu.pumping}. As this layer is also where rotational
shear is maximal, it is plausible that the strong toroidal fields are not just
stored but also generated here, by the winding up of poloidal field. The two
main groups of dynamo models, interface dynamos and flux transport dynamos,
differ mainly in their assumptions about the site and mechanism of the 
$\alpha$-effect responsible for the generation of a new poloidal field from the
toroidal field. 

In interface dynamos $\alpha$ is assumed to be concentrated near the bottom of
the convective zone, in a region adjacent to the tachocline, so that the dynamo
operates as a wave propagating along the interface between these two layers.
While these models may be roughly consistent and convincing from the
physical point of view, they have only had limited success in reproducing the
observed characteristics of the solar cycle, such as the butterfly diagram. 

Flux transport dynamos, in contrast, rely on the Babcock--Leighton mechanism
for $\alpha$, arising due to the action of the Coriolis force on emerging flux
loops, and they assume that the corresponding $\alpha$-effect is concentrated
near the surface. They keep this surface layer \textit{incommunicado} with the
tachocline by introducing some arbitrary unphysical assumptions (such as very
low diffusivities in the bulk of the convective zone). The poloidal fields
generated by this surface $\alpha$-effect are then advected to the poles and
there down to the tachocline by the meridional circulation -- which,
accordingly, has key importance in these models. The equatorward deep return
flow of the meridional circulation is assumed to have a significant overlap
with the tachocline (another controversial point), and it keeps transporting
the toroidal field generated by the rotational shear towards the equator. By
the time it reaches lower latitudes, it is amplified sufficiently for the flux
emergence process to start, resulting in the formation of active regions and,
as a result of the Babcock--Leighton mechanism, in the reconstruction of a
poloidal field near the surface with a polarity opposed to that in the previous
11-year cycle. While flux transport models may be questionable from the point
of view of their physical consistency, they can be readily fine-tuned to
reproduce the observed butterfly diagram quite well.

It should be noted that while the terms ``interface dynamo'' and ``flux
transport dynamo'' are now very widely used to describe the two main approaches,
the more generic terms ``advection-dominated'' and ``diffusion-dominated'' would
be preferable in several respects. This classification allows for a
continuous spectrum of models depending on the numerical ratio of advective and
diffusive timescales (for communication between surface and tachocline). In
addition, even at the two extremes, classic interface dynamos and
circulation-driven dynamos are just particular examples of advection or
diffusion dominated systems with different geometrical structures.

\subsection{Is model-based cycle prediction feasible?}

As it can be seen even from the very brief and sketchy presentation given above,
all current solar dynamo models are based on a number of quite arbitrary
assumptions and depend on a number of free parameters, the functional form and
amplitude of which is far from being well constrained. For this reason, 
\cite{Bushby+Tobias} rightfully say that all current solar dynamo models are
only of ``an illustrative nature''. This would suggest that as far as solar
cycle prediction is concerned, the best we should expect from dynamo models is
also an ``illustrative'' reproduction of a series of solar cycles with the same
kind of long-term variations (qualitatively and, in the statistical sense,
quantitatively) as seen in solar data. Indeed, \cite{Bushby+Tobias} demonstrated
that even a minuscule stochastic variation in the parameters of a particular
flux transport model can lead to large, unpredictable variations in the cycle
amplitudes. And even in the absence of stochastic effects, the chaotic nature of
nonlinear dynamo solutions seriously limits the possibilities of prediction, as
the authors find in a particular interface dynamo model: even if the very same
model is used to reproduce the results of one particular run, the impossibility
of setting initial conditions exactly representing the system implies that
predictions are impossible even for the next cycle. Somewhat better results are
achieved by an alternative method, based on the phase space reconstruction of
the attractor of the nonlinear system -- this is, however, a purely empirical
time series analysis technique for which no knowledge of the detailed underlying
physics is needed. (Cf.\ Section~\ref{sect:nonlin} above.)

Despite these very legitimate doubts regarding the feasibility of model-based
prediction of solar cycles, in recent years several groups have claimed to be
able to predict solar cycle 24 on the basis of dynamo models with a high
confidence. So let us consider these claims.

\subsection{Explicit models}
\label{sect:explicit}

The current buzz in the field of model-based solar cycle prediction was started
by the work of the solar dynamo group in Boulder \citep{Dikpati:forecast1st, 
Dikpati+:prediction}. Their model is a flux transport dynamo,
advection-dominated to the extreme. The strong suppression of diffusive effects
is assured by the very low value (less than $20\,$km$^2/$s) assumed for the
turbulent magnetic diffusivity in the bulk of the convective zone. As a result,
the poloidal fields generated near the surface by the Babcock--Leighton
mechanism are only transported to the tachocline on the very long, decadal time
scale of meridional circulation. The strong toroidal flux residing in the
low-latitude tachocline, producing solar activity in a given cycle is thus the
product of the shear amplification of poloidal fields formed near the surface
about 2--3 solar cycles earlier, i.e. the model has a ``memory'' extending to
several cycles. The mechanism responsible for cycle-to-cycle variation is
assumed to be the stochastic nature of the flux emergence process. In order to
represent this variability realistically, the model drops the surface
$\alpha$-term completely (a separate, smaller $\alpha$ term is retained in the
tachocline); instead, the generation of poloidal field near the
surface is represented by a source term, the amplitude of which is based on the
sunspot record, while its detailed functional form remains fixed. 

\cite{Dikpati+:prediction} find that, starting off their calculation by fixing
the source term amplitudes of sunspot cycles~12 to 15, they can predict the
amplitudes of each subsequent cycle with a reasonable accuracy, provided that
the relation between the relative sunspot numbers and the toroidal flux in the
tachocline is linear, and that the observed amplitudes of all previous cycles
are incorporated in the source term for the prediction of any given cycle. For
the upcoming cycle~24 the model predicts peak smoothed annual relative sunspot
numbers of 150 or more. Elaborating on their model, they proceeded to apply it
separately to the northern and southern hemispheres, to find that the model can
also be used to correctly forecast the hemispheric asymmetry of solar activity
\citep{Dikpati+:hemisph.prediction}.

The extraordinary claims of this pioneering research have prompted a hot debate
in the dynamo community. Besides the more general, fundamental doubt regarding
the feasibility of model-based predictions (see Section~3.2 above),
more technical concerns arose, to be discussed below. 

Another flux transport dynamo code, the Surya code, originally developed by A.
Choudhuri and coworkers in Bangalore, has also been utilized for prediction
purposes. The crucial difference between the two models is in the value of the
turbulent diffusivity assumed in the convective zone: in the Bangalore model
this value is $240\,$km$^2/$s, 1--2 orders of magnitude higher than in the
Boulder model, and within the physically plausible range
\citep{Chatterjee+:model}. As a result of the shorter diffusive timescale, the
model has a shorter memory, not exceeding one solar cycle. As a consequence of
this relatively rapid diffusive communication between surface and tachocline,
the poloidal fields forming near the surface at low latitudes due to the
Babcock--Leighton mechanism diffuse down to the tachocline in about the same
time as they reach the poles due to the advection by the meridional circulation.
In these models, then, polar magnetic fields are not a true physical precursor
of the low-latitude toroidal flux, and their correlation is just due to their
common source. In the version of the code adapted for cycle prediction
\citep{Choudhuri+:prediction, Jiang+:prediction}, the ``surface'' poloidal field
(i.e. the poloidal field throughout the outer half of the convection zone) is
rescaled at each minimum by a factor reflecting the observed amplitude of the
Sun's dipole field. The model shows reasonable predictive skill for the last
three cycles for which data are available, and can even tackle hemispheric
asymmetry \citep{Goel+:hemispheric}. For cycle 24, the predicted amplitude is
30--35\% lower than cycle~23.

\subsection{Truncated models}
\label{sect:truncated}

The ``illustrative'' nature of solar dynamo models is nowhere more clearly on
display than in truncated or reduced models where some or all of the detailed
spatial structure of the system is completely disregarded, and only temporal
variations are explicitly considered. This is sometimes rationalised as a
truncation or spatial integration of the equations of a more realistic
inhomogenous system; in other cases, no such rationalisation is provided,
representing the solar dynamo by an infinite, homogeneous or periodic turbulent
medium where the amplitude of the periodic large-scale magnetic field varies
with time only.

In the present subsection we deal with models that do keep one spatial variable
(typically, the latitude), so growing wave solutions are still possible
-- these models, then, are still dynamos even though their spatial
structure is not in a good correspondence with that of the solar dynamo.

This approach in fact goes back to the classic migratory dynamo model of
\cite{Parker:migr.dynamo} who radially truncated (i.e., integrated) his
equations to simplify the problem. Parker seems to have been the first to
employ a heuristic relaxation term of the form $-B_r/\tau_d$ in the poloidal
field equation to represent the effect of radial diffusion; here,
$\tau_d=d^2/\eta_T$ is the diffusive timescale across the thickness $d$ of the
convective zone. His model  was recently generalized by
\cite{Moss+:stochastic.dynamo1} and \cite{Moss+:stochastic.dynamo2} to the case
when the $\alpha$-effect includes an additive stochastic noise, and nonlinear
saturation of the dynamo is achieved by $\alpha$-quenching. These authors do
not make an attempt to predict solar activity with their model but they can
reasonably well reproduce some features of the very long term solar activity
record, as seen from cosmogenic isotope studies.

Another radially truncated model, this time formulated in a Cartesian system, is
that of \cite{Kitiashvili+:nonlin.dynamo}. In this model stochastic effects are
not considered, and, in addition to using an $\alpha$-quenching recipe, further
nonlinearity is introduced by coupling in the Kleeorin--Ruzmaikin equation
\citep{Zeldovich+:mg.book} governing the evolution of magnetic helicity, which
in the hydromagnetic case contributes to $\alpha$. Converting the toroidal field
strength to relative sunspot number using the Bracewell transform,
eq.~(\ref{eq:Bracewell}), the solutions reproduce the asymmetric profile of the
sunspot number cycle. For sufficiently high dynamo numbers the solutions become
chaotic, cycle amplitudes show an irregular variation. Cycle amplitudes and
minimum--maximum time delays are found to be related in a way reminiscent of the
Waldmeier relation. 

Building on these results, \cite{Kitiashvili+:prediction} attempt to predict
solar cycles using a data assimilation method. The approach used is the
so-called Ensemble Kalman Filter method. Applying the model for a
``postdiction'' of the last 8 solar cycles yielded astonishingly good results,
considering the truncated and arbitrary nature of the model and the fundamental
obstacles in the way of reliable prediction discussed above. While the presently
available brief preliminary publication leaves several details of the method
unclear, the question may arise whether the actual physics of the model
considered has any significant role in this prediction, or we are dealing with
something like the phase space reconstruction approach discussed in
Section~\ref{sect:nonlin}  above where basically any model with an attractor
that looks reasonably similar to that of the actual solar dynamo would do.
Either way, the method is remarkable, and the prediction for cycle~24 of a
maximal smoothed annual sunspot number of 80, to be reached in 2013, will be
worth comparing to the actual value.

In order to understand the origin of the predictive skill of the Boulder model,
\cite{Cameron+:prediction} studied a radially truncated version of the model,
wherein only the equation for the radial field component is solved as a function
of time and latitude. The equation includes a source term similar to the one
used in the Boulder model. As the toroidal flux does not figure in this simple
model, the authors use the transequatorial flux $\Phi$ as a proxy, arguing that
this may be more closely linked to the amplitude of the toroidal field in the 
upcoming cycle than the polar field. They find that $\Phi$ indeed correlates
quite well (correlation coefficients $r\sim0.8$--$0.9$, depending on model
details) with the amplitude of the next cycle, as long as the form of the
latitude dependence of the source term is prescribed and only its amplitude is
modulated with the observed sunspot number series (``idealized model''). But
surprisingly, the predictive skill of the model is completely lost if the
prescribed form of the source function is dropped and the actually observed
latitude distribution of sunspots is used instead (``realistic model'').
\cite{Cameron+:prediction} interpret this by pointing out that $\Phi$ is mainly
determined by the amount of very low latitude flux emergence, which in turn
occurs mainly in the last few years of the cycle in the idealized model, while
it has a wider temporal distribution in the realistic model. The conclusion is
that the root of the apparently good predictive skill of the truncated model
(and, by inference, of the Boulder model it is purported to represent) is
actually just the good empirical correlation between late-phase activity and the
amplitude of the next cycle, discussed in Section~\ref{sect:minimax} above. This
correlation is implicitly ``imported'' into the idealized flux transport model
by assuming that the late-phase activity is concentrated at low latitudes, and
therefore gives rise to cross-equatorial flux which then serves as a seed for
the toroidal field in the next cycle. So if \cite{Cameron+:prediction} are
correct, the predictive skill of the Boulder model is due to an empirical
precursor and is thus ultimately explained by the good old Waldmeier effect (cf.
Section~\ref{sect:Waldmeier})

The fact that the truncated model of \cite{Cameron+:prediction} is not
identical to the Boulder model obviously leaves room for doubt regarding this
conclusion. In particular, the effective diffusivity represented by the sink
term in the truncated model is $\sim 100\,$km$^2/$s, significantly higher than
in the Boulder model; consequently, the truncated model will have a more
limited memory, cf.\ \cite{Yeates+:prediction}. The argument that the
cross-equatorial flux is a valid proxy of the amplitude of the next cycle may
be correct in such a short-memory model with no radial structure, but it is
dubious whether it remains valid for flux transport models in general. In an
attempt to appreciate the importance of the cross-equatorial flux in their
model, \cite{Dikpati+:crosseq} find that while this flux does indeed correlate
fairly well ($r=0.76$) with the next cycle amplitude, the toroidal flux is a
much better predictor ($r=0.96$). At first sight this seems to make it unlikely
that the former can explain the latter; however, part of the difference in the
predictive skill may be due to the fact that $\Phi$ shows much more short-term
variability than the toroidal flux. 

In any case, the obvious way to address the concerns raised by
\cite{Cameron+:prediction} and further by \cite{Schussler:prediction} in
relation to the Boulder model would be to run that model with a modified source
function incorporating the realistic latitudinal distribution of sunspots in
each cycle. The results of such a test are not yet available at the time of
writing this review.

\subsection{The Sun as an oscillator}
\label{sect:oscillator}

An even more radical simplification of the solar dynamo problem ignores any
spatial dependence in the solutions completely, concentrating on the time
dependence only. Spatial derivatives appearing in Equations~(\ref{eq:pol}) and
(\ref{eq:tor}) are estimated as $\nabla\sim 1/L$ and the resulting terms $U_c/L$
and $\eta_T/L^2$ as $1/\tau$ where $\tau$ is a characteristic time scale. This
results in the pair 
\begin{equation} 
  \dot A=\alpha B -A/\tau
  \label{eq:A2} 
\end{equation}
\begin{equation} 
  \dot B=({\Omega}/L) A -B/\tau
  \label{eq:B2} 
\end{equation} 
which can be combined to yield 
\begin{equation}
  \ddot B=\frac{D-1}{\tau^2} B -\frac 2\tau\dot B
  \label{eq:sunosc}
\end{equation} 
where $D=\alpha\Omega\tau^2/L$ is the dynamo number. For $D<1$,
Equation~(\ref{eq:sunosc}) clearly describes a damped linear
oscillator. For $D>1$, solutions have a non-oscillatory character. The
system described by Equation~(\ref{eq:sunosc}), then, is not only not
a true dynamo (missing the spatial dependence), but it does not even
display growing oscillatory solutions that would be the closest
counterpart of dynamo-like behaviour in such a system. Nevertheless,
there are a number of ways to extend the oscillator model to allow for
persistent oscillatory solutions, i.e., to turn it into a
\textit{relaxation oscillator.}

(1) The most straightforward approach is to add a \textit{forcing term}
$+\sin(\omega_0t)$ to the r.h.s.\ of Equations~(\ref{eq:sunosc}). Damping would cause
the system to relax to the driving period $2\pi/\omega_0$ if there were no
stochastic disturbances to this equilibrium. \cite{Hiremath:oscillator}
fitted the parameters of the forced and damped oscillator model to each observed
solar cycle individually; then in a later work \citep{Hiremath} he applied
linear regression to the resulting series to provide a forecast (see
Section~\ref{sect:regression} above).

(2) Another trick is to account for the $\pi/2$ phase difference between
poloidal and toroidal field components in a dynamo wave by introducing a {\it
phase factor} $i$ into the first term on the r.h.s.\ of Equation~(\ref{eq:B2}).
This can also be given a more formal derivation as equations of this form result
from the substitution of solutions of the form $A\propto e^{ikx}$, $B\propto
e^{i(kx+\pi/2)}$ into the 1D dynamo equations. This route, combined with a
nonlinearity due to magnetic modulation of differential rotation described by a
coupled third equation, was taken by \cite{Weiss+:chaotic.dynamo}. Their model
displayed chaotic behaviour with intermittent episodes of low activity similar
to grand minima. 

(3) \cite{WilmotSmith+} showed that another case where dynamo-like behaviour can
be found in an equation like~(\ref{eq:sunosc}) is if the missing effects of
finite communication time between parts of a spatially extended system are
reintroduced by using a \textit{time delay} $\Delta t$, evaluating the first
term on the r.h.s.\ at time $t-\Delta t$ to get the value for the l.h.s.\ at
time $t$.

(4) Yet another possibility is to introduce a \textit{nonlinearity} into the model
by assuming $D=D_0[1-f(B)]$ where $f(B=0)=0$ and $f\geq 0$ everywhere. (Note
that any arbitrary form of $\alpha$- or $\Omega$-quenching can be cast in the
above form by series expansion.) The governing equation then becomes one of a
nonlinear oscillator: 
\begin{equation} 
\ddot B=\frac{D_0-1}{\tau^2} B -\frac 2\tau\dot B-\frac{D_0-1}{\tau^2} B f(B)  
  \label{eq:sunosc2}
\end{equation} 

In the most commonly assumed quenching mechanisms the leading term in $f(B)$ is
quadratic; in this case Equation~(\ref{eq:sunosc2}) describes a \textit{Duffing
oscillator} \citep{Kanamaru:Duffing}. For large positive dynamo numbers,
$D_0\gg 1$, then, the large nonlinear term dominates for high values of $B$, its
negative sign imposing oscillatory behaviour; yet the origin is a repeller so
the oscillation will never be damped out.  The Duffing oscillator was first
considered in the solar context by \cite{Palus+Novotna}. Under certain
conditions on the parameters, it can be reduced to a \textit{van der Pol
oscillator} \citep{Adomian, Mininni+:vanderpol2, Kanamaru:vdPol}:
\begin{equation} 
  \ddot \xi= -\xi +\mu(1-\xi^2)\dot\xi \,,
  \label{eq:vanderPol}
\end{equation} 
with $\mu>0$. From this form it is evident that the problem is equivalent to
that of an oscillator with a damping that increases with amplitude; in fact, for
small amplitudes the damping is negative, i.e., the oscillation is self-excited.

These simple nonlinear oscillators were among the first physical systems where
chaotic behaviour was detected (when a periodic forcing was added). Yet,
curiously, they first emerged in the solar context precisely as an alternative
to chaotic behaviour. Considering the mapping of the solar cycle in the
differential phase space $\{B, dB/dt\}$, \cite{Mininni+:vanderpol} got the
impression that, rather than showing signs of a strange attractor. the SSN
series is adequately modelled by a van der Pol oscillator with stochastic
fluctuations.  This concept was further developed by \cite{Lopes+Passos:dalton}
who fitted the parameters of the oscillator to each individual sunspot cycle.
The parameter $\mu$ is related to the meridional flow speed and the fit
indicates that a slower meridional flow may have been responsible for the Dalton
minimum. This was also corroborated in an explicit dynamo model (the Surya code)
---however, as we discussed in Sect.~\ref{sect:flows}, this result of flux
transport dynamo models is spurious and the actual effect of a slower meridional
flow is likely to be opposite to that suggested by the van der Pol oscillator
model.

In an alternative approach to the problem, \cite{Nagovitsyn:reconstr} attempted
to constrain the properties of the solar oscillator from its
amplitude--frequency diagram, suggesting a Duffing oscillator driven at two
secular periods. While his empirical reconstruction of the amplitude--frequency
plot may be subject to many uncertainties, the basic idea is certainly
noteworthy.

In summary: despite its simplicity, the oscillator representation of the solar
cycle is a relatively new development in dynamo theory, and its obvious
potential for forecasting purposes has barely been exploited.


\newpage

\section{Summary Evaluation}
\label{sec:Summary-Evaluation}

The performance of various forecast methods in cycles 21--23 was discussed by
\cite{Li+:pred.rev} and \cite{Kane:pred23rev}. 

\textit{Precursor methods} stand out with their internally consistent forecasts for
these cycles which for cycles 21 and 22 proved to be correct. For cycle~23 these
methods were still internally consistent in their prediction, mostly scattering
in a narrow range between 150 and 170; however, the cycle amplitude proved to be
considerably lower ($\Rmax=121$). It should be noted, however, that one
precursor based prediction, that of \cite{Schatten+:pred23} was significantly
lower than the rest ($138\pm 30$) and within $0.6\sigma$ of the actual value.
Indeed the method of Schatten et al.\ (\citeyear{Schatten+:pred22,
  Schatten+:pred23}), has consistently proven its skill in all
cycles. As discussed in Sect.~\ref{sect:polar}, this method is
essentially based on the polar magnetic field strength as precursor.

\textit{Extrapolation methods} as a whole have shown a much less impressive
performance. Overall, the statistical distribution of maximum amplitude values
predicted by ``real'' forecasts made using these methods (i.e.\ forecasts made
at or before the minimum epoch) for any given cycle does not seem to
significantly differ from the long term climatological average of the solar
cycle quoted in Section~\ref{sect:cycle} above ($100\pm 35$). It would of course
be a hasty judgement to dismiss each of the widely differing individual
approaches comprised in this class simply due to the poor overall performance of
the group. In particular, some novel methods suggested in the last 20 years,
such as SSA or neural networks have hardly had a chance to debut, so their
further performance will be worth monitoring in upcoming cycles.

One group of extrapolation methods that stands apart from the rest are those
based on the even--odd rule. These methods enjoyed a relatively high prestige
until cycle~23, when they coherently predicted a peak amplitude around 200,
i.e., $\sim70\%$ higher than the actual peak. This can only be qualified as a
miserable failure, independently of the debate as to whether cycle~23 is
truly at odds with the even--odd rule or not.

In this context it may be worth noting that the double peaked character and long
duration of cycle~23 implies that its \textit{integrated} amplitude (sum of annual
sunspot numbers during the cycle) is much less below that of cycle~22 than the
peak amplitude alone would indicate. This suggests that forecasts of the
integrated amplitude (rarely attempted) could be more robust than forecasts of
the peak. Nevertheless, one has to live with the fact that for most practical
applications (space weather) it is the peak amplitude that matters most, so this
is where the interest of forecasters is naturally focused.

Finally, \textit{model based methods} are a new development that have had no
occasion yet to prove their skill. As discussed above, current dynamo models do
not seem to be at a stage of development where such forecasts could be attempted
with any confidence, especially before the time of the minimum. (The method of
\citealp{Choudhuri+:prediction}, using polar fields as input near the minimum,
would seem to be akin to a version of the polar field based precursor method
with some extra machinery built into it.) The claimed good prediction skills of
models based on data assimilation will need to be tested in future cycles and
the roots of their apparent success need to be understood.

Table~\ref{tab:1} presents a collection of forecasts for the amplitude of cycle~24,
without claiming completeness. \citep[See, e.g.,][for a more exhaustive
list.]{Pesnell} The objective was to include one or two representative forecasts from
each category. 

\begin{table}
\caption{A selection of forecasts for cycle 24.}
\label{tab:1}
\begin{tabular}{lrll}
\toprule
Category & Peak amplitude & Link & Reference \\
\midrule
Precursor methods\\
\bek Minimax & $80 \pm 25$ &  Eq.~\ref{eq:minimax} & \cite{Brown:minimax}; \cite{Brajsa+:gleissbg}$^*$\\
\bek Minimax3 &  $69 \pm 15$ & Eq.~\ref{eq:minimax3} &
\cite{Cameron+:prediction}$^*$\\
\bek Polar field & $75 \pm 8$ & Sect.~\ref{sect:polar} & \cite{Svalgaard+:prediction24} \\
\bek Polar field & $80 \pm 30$ & Sect.~\ref{sect:polar} & \cite{Schatten+:pred24}\\
\bek Geomagnetic (Feynman) & $150$ & Sect.~\ref{sect:geomg} & \cite{Hathaway+Wilson}\\
\bek Geomagnetic (Ohl) & $93\pm 20$ & Sect.~\ref{sect:geomg} & \cite{Bhatt+}\\
\bek Geomagnetic (Ohl) & $101\pm 5$ & Sect.~\ref{sect:evenodd} & \cite{Ahluwalia:pred24}\\
\bek Geomagnetic (interpl.)& $97\pm 25$ & Sect.~\ref{sect:geomg} & \cite{Wang+Sheeley:geomg.precursor}\\
\bek Field reversal & $94\pm 14$& Eq.~\ref{eq:tlatov}
 & \cite{Tlatov:polar.precursors}$^*$\\
Extrapolation methods\\
\bek Linear regression & $90\pm 27$ & Sect.~\ref{sect:regression} & \cite{Brajsa+:gleissbg}\\
\bek Linear regression & $110\pm 10$ & Sect.~\ref{sect:regression} & \cite{Hiremath}\\
\bek Spectral (MEM) & $90\pm 11$ & Sect.~\ref{sect:spectral} & \cite{Kane:MEM24}\\
\bek Spectral (SSA) & $117$ & Sect.~\ref{sect:spectral} & \cite{Loskutov:SSA}\\
\bek Spectral (SSA) & $106$ & Sect.~\ref{sect:spectral} & \cite{Kuzanyan:predict.poster}\\
\bek Attractor analysis & $87$ & Sect.~\ref{sect:nldpro} & \cite{Kilcik+}\\
\bek Attractor analysis & $65\pm 16$ & Sect.~\ref{sect:nldpro} & \cite{Aguirre+}\\
\bek Attractor analysis & $145\pm 7$ & Sect.~\ref{sect:nldpro} & \cite{Crosson+Binder}\\
\bek Neural network & $145$ & Sect.~\ref{sect:neural} & \cite{Maris+Oncica}\\
\bek Neural network & $117.5\pm 8.5$ & Sect.~\ref{sect:neural} & \cite{Uwamahoro}\\
Model based methods\\
\bek Explicit models & $167\pm 12$ & Sect.~\ref{sect:explicit} & \cite{Dikpati+:prediction}\\
\bek Explicit models & $\sim 80$ & Sect.~\ref{sect:explicit} & \cite{Choudhuri+:prediction}\\
\bek Explicit models & $\sim 85$ & Sect.~\ref{sect:explicit} & \cite{Jiang+:prediction}\\
\bek Truncated models & $\sim 80$ & Sect.~\ref{sect:truncated} & \cite{Kitiashvili+:prediction}\\
\bottomrule
\end{tabular}
{\footnotesize References marked with $^*$ are to the basic principle used in
the given prediction method while the actual numerical evaluation for cycle~24
was done by the author. The application for forecast purposes does not
necessarily reflect the original intention of the basic principle, as laid out
in the cited publications.}
\end{table}

The incipient cycle~24 may be a milestone for solar cycle forecasting. Current
evidence indicates that we are at the end of the Modern Maximum when the Sun is
about to switch to a state of less intense long term activity. The appearance
of a number of novel prediction methods, in particular the model based approach,
as well as the unusually large discrepancy between forecasts based on the
precursor approach imply that, whichever course solar activity will take in the
coming years, we have a lot to learn from the experience.


\newpage

\section{Epilogue}
\label{sec:Epilogue}

Throughout the ages, mankind felt and tried to answer the urge to predict events
to  come. \textit{Omens} were carefully collected and categorized on Mesopotamian
clay tablets; omen-based prediction was developed into an industry in the form
of hepatoscopy (analyzing the shape of the liver of a sacrificed animal) and, in
later Roman times, of auspicium (watching the flight of the birds). Ancient
Greeks often turned to \textit{oracles} like the Pythia of Delphi. By the late
Antiquity, the astrological world view was widespread throughout the civilized
world, implying that cosmic and terrestrial events were subject to \textit{cosmic
cycles} governed by a variety of (planetary) periods. 

Today we tend to smile at these ``superstitious'' early attempts. Yet,
ironically, many of the ``advanced'' methods we have for the prediction of solar
activity are based on principles that hardly differ from those listed above:
just substitute ``precursor'' for ``omen'', ``neural network'' for ``oracle'' or
``harmonic analysis'' for ``cosmic cycles''...

But in parallel with the often na{\"\i}ve phenomenological or empirical
prediction attempts, already in the Hellenistic world, a handful of enlightened
scientists started the development of physical models, based on logic and
experience, that would lead to the advanced predictive skills of many
models of modern science \citep{Russo}. Extending the analogy, we can see that
the real importance of the recent debut of model-based solar cycles predictions
is not their still dubious success rate but the conceptual leap they represent. 

Despite the rather poor overall performance of solar cycle prediction attempts,
the extensive efforts invested in this endeavour were not in vain as they have
contributed and keep contributing to a better understanding of the physical
processes governing the solar cycle and to constraining the dynamo.



\section{Acknowledgements}
\label{section:acknowledgements}

Support by the Hungarian Science Research Fund (OTKA grant no.\ K67746), by the
European Commission's 6th Framework Programme (SOLAIRE Network,
MTRN-CT-2006-035484) as well as by the European Union with the co-financing of
the European Social Fund (grant no.\ T\'AMOP-4.2.1/B- 09/1/KMR-2010-0003) is
gratefully acknowledged.

Wilcox Solar Observatory data used in this study was obtained via the web site
\url{http://wso.stanford.edu/}, courtesy of J.T.~Hoeksema.
The Wilcox Solar Observatory is currently supported by NASA.


\newpage


\bibliography{refs}

\end{document}